\documentclass[a4paper,11pt,preprintnumbers]{article}
\pdfoutput=1 

\usepackage{jcappub} 

\usepackage{mathabx}
\usepackage{mathtools}
\usepackage{appendix}
\usepackage{slashed}
\usepackage{multirow}
\usepackage{array}
\usepackage{cellspace}
\usepackage{xcolor}
\usepackage{makecell}
\usepackage{braket}
\usepackage{amsmath, amsfonts, amssymb}
\usepackage{mathrsfs}
\usepackage{subcaption}
\usepackage[normalem]{ulem}

\newcommand{\Tr}{{\rm Tr}}

\title{\boldmath Cosmic neutrino background detection in the minimally extended Standard Model}

\author[1,2]{Yuber F. Perez-Gonzalez}
\author[3]{and Jack D. Shergold}

\affiliation[1]{Institute for Particle Physics Phenomenology, Durham University,\\ South Road, DH1 3LE, Durham, United Kingdom}
\affiliation[2]{Departamento de Física Teórica and Instituto de Física Teórica UAM/CSIC,\\
Universidad Autónoma de Madrid, Cantoblanco, 28049 Madrid, Spain}
\affiliation[3]{Instituto de Física Corpuscular (IFIC), CSIC,\\Parc Científic, C/Catedrático José Beltrán,
2, E-46980,\\Paterna, Spain}

\preprintnumber{\newline IPPP/24/68, IFT-UAM/CSIC-24-146}

\emailAdd{yuber.perez@uam.es}
\emailAdd{jack.d.shergold@ific.uv.es}

\abstract{    
    We investigate the sensitivity of relic neutrino detection methods within the Standard Model, extended to include right-chiral neutrino singlets with Majorana mass terms. In particular, we study neutrino capture on unstable nuclei, the Stodolsky effect, coherent scattering, and an accelerator experiment. We demonstrate that the sensitivity transitions smoothly between Dirac and Majorana regimes, depending on the scale of lepton number violation. Importantly, neutral current interactions lead to transitions between the light and heavy neutrino states, necessitating the use of a density matrix formalism for accurate sensitivity calculations.
    As the oldest source of neutrinos in the universe, relic neutrinos would be able to provide an ultimate constraint on the lepton number violating scale, $m_R\gtrsim 10^{-33}~{\rm eV}$, below which neutrinos would behave as Dirac fermions for all practical purposes.
}

\begin{document}
\maketitle

\flushbottom
\def\thefootnote{\arabic{footnote}}
\setcounter{footnote}{0}

\section{Introduction}\label{sec:introduction}
The cosmic neutrino background (C$\nu$B) is a firm prediction of the $\Lambda$CDM model of cosmology, a sea of primordial neutrinos which today remain undetected due to the combination of their exclusively weak interactions and sub-meV temperature. A successful detection of the C$\nu$B, produced approximately one second after the Big Bang, could give key insight into the evolution of the universe. In particular, as the C$\nu$B predates the first atomic nuclei, its detection would give us a window into the processes that gave rise to the element abundances observed today. Additionally, as a consequence of its low temperature, many experiments aiming to detect the C$\nu$B are also sensitive to both the neutrino mass, as well as the Dirac or Majorana nature of neutrinos~\cite{Long:2014zva,Bauer:2022lri}.

Despite the extreme challenges involved, there are many promising proposals to detect the C$\nu$B: thresholdless capture of neutrinos on radioactive nuclei~\cite{Long:2014zva,Weinberg:1962zza,Cocco:2007za,Cocco:2009rh,Betts:2013uya,Akhmedov:2019oxm,PTOLEMY:2022ldz}; coherent scattering of relic neutrinos on macroscopic targets~\cite{Opher:1974drq,Lewis:1979mu,Shvartsman:1982sn,Cabibbo:1982bb,Smith:1983jj,Ringwald:2004np,Duda:2001hd,Gelmini:2004hg,Domcke:2017aqj,Shergold:2021evs}; spin-dependent energy shifts of Standard Model (SM) fermions in a background of relic neutrinos~\cite{Stodolsky:1974aq,Duda:2001hd,Gelmini:2004hg,Domcke:2017aqj,Rostagni:2023eic}; ultrahigh energy cosmic neutrino scattering on the C$\nu$B at the $Z$-resonance~\cite{Eberle:2004ua,Ringwald:2004np}; capturing neutrinos on an ultrarelativistic ion beam~\cite{Bauer:2021uyj}; modifications to atomic de-exicitation rates due to Pauli blocking~\cite{Yoshimura:2014hfa}. These have all been reviewed alongside current experimental, theoretical and cosmological constraints on the C$\nu$B in~\cite{Bauer:2022lri}. More recently, there have also been proposals to detect the C$\nu$B using a diffraction grating~\cite{Arvanitaki:2023fij}, observing loop induced bremsstrahlung from the scattering of relic and solar neutrinos~\cite{Asteriadis:2022zmo}, and indirectly through the anomalous cooling of neutron stars~\cite{Das:2024thc, Chauhan:2024deu}.

Another open question is the origin of neutrino mass, shown extensively by neutrino oscillation experiments to be non-zero~\cite{Super-Kamiokande:1998kpq,SNO:2001kpb,KamLAND:2002uet,MINOS:2006foh,T2K:2011ypd}. In the minimally extended Standard Model (MESM), neutrino mass is generated by the addition of right-chiral, sterile counterparts to the left-chiral SM neutrino fields. This in turn allows us to write down a Dirac mass term that mixes the active and sterile neutrinos, akin to those of the other SM fermions, along with a Majorana mass term for the sterile components. When the Majorana mass far exceeds the Dirac mass, the light and heavy mass eigenstates remain distinct, and approximately align with the active and sterile states, respectively. This is the familiar seesaw mechanism, typically used to explain the smallness of the active neutrino mass~\cite{Schechter:1980gr, Schechter:1981cv}. As we will show in Section~\ref{sec:mesm}, if instead the Dirac mass dominates over the Majorana mass, the mass eigenstates states mix maximally, leading to oscillations between the active and sterile neutrino states. This is the pseudo-Dirac neutrino. Once the Majorana mass becomes sufficiently small, the active-sterile oscillation baseline becomes so large that pseudo-Dirac neutrinos become indistinguishable from Dirac neutrinos, and all lepton number violating processes proceed at unobservable rates. As the oldest source of free-streaming neutrinos, the C$\nu$B is uniquely sensitive to tiny Majorana masses that give rise to oscillations with periods of order the age of the universe.

In this paper, we will explore the sensitivity of a variety of relic neutrino detection proposals in the MESM. The effect of pseudo-Dirac neutrinos on signals at the PTOLEMY experiment has already been explored in~\cite{Perez-Gonzalez:2023llw}, where the authors demonstrate that for some combinations of parameters, the pseudo-Dirac capture rate is lower than that of purely Dirac neutrinos. Here we will extend this analysis to the Stodolsky effect~\cite{Stodolsky:1974aq,Duda:2001hd,Gelmini:2004hg,Domcke:2017aqj,Rostagni:2023eic}, coherent scattering~\cite{Opher:1974drq,Lewis:1979mu,Shvartsman:1982sn,Cabibbo:1982bb,Smith:1983jj,Ringwald:2004np,Duda:2001hd,Gelmini:2004hg,Domcke:2017aqj,Shergold:2021evs}, and accelerator proposals~\cite{Bauer:2021uyj}, and demonstrate the smooth transition in the sensitivities from the seesaw limit to the Dirac limit discussed in~\cite{Bauer:2022lri}, as the Majorana mass tends to zero. In the process, we will also find the smallest Majorana masses that can be distinguished at a future relic neutrino detection experiment.

The remainder of this paper will be organised as follows. In Section~\ref{sec:mesm} we will introduce the MESM, and discuss our procedure for computing neutrino phases. Next, in Section~\ref{sec:detectors} we will introduce several C$\nu$B detection proposals, and give expressions for their sensitivity to the C$\nu$B in the MESM. Following this, we will show the sensitivity of each proposal to relic neutrinos over a broad Dirac and Majorana mass parameter space in Section~\ref{sec:results}, before concluding in Section~\ref{sec:conclusions}.

\section{Minimally extended Standard Model}\label{sec:mesm}

The observed $L/E$ dependence of neutrino oscillations, with $L$ the source-detector distance and $E$ the neutrino energy, has firmly established that neutrino oscillations originate from the mismatch of the mass and flavour bases. This demonstrates that neutrinos are massive, and thus the existence of beyond Standard Model physics. When trying to extend the SM to include neutrino masses, we encounter a unique situation. As the combination $\overline{L}_\alpha \widetilde{H}$ between left-chiral lepton doublets, $L_\alpha$, and the conjugate of the SM Higgs doublet, $\widetilde{H}$, has zero total hypercharge, we would need to introduce right-chiral fermionic singlets $\nu_{i,R}$, with $i=\{1,2,3\}$, to build neutrino mass terms. However, as nothing forbids the existence of Majorana mass terms for such singlets, the most general renormalisable Lagrangian for neutrino masses is
\begin{align}\label{eq:mass_lagr_1}
    \mathcal{L}_{\nu}  = -Y_{\alpha i} \overline{L}_\alpha \widetilde{H} \nu_{i,R} + \frac{1}{2}\overline{(\nu_{i,R})^c} (M_R^*)_{ij} \nu_{j,R}+{\rm h.c.}\, ,
\end{align}
where $\alpha \in \{e,\mu,\tau\}$, $Y$ denotes the matrix of Yukawa couplings, and $M_R$ denotes the matrix of Majorana masses for the singlets\footnote{$M_R$ is defined with a complex conjugation to make diagonalisation more convenient. See~\cite{Kobayashi:2000md} for more details.}. We will refer to this scenario as the minimally extended Standard Model (MESM) in what follows.

As the mass terms for $\nu_{i,R}$ are not protected by the electroweak gauge symmetry, the Majorana mass scale could be much higher than the electroweak scale, reaching up to the Grand Unification scale, or conversely, could be much lower than the electroweak scale. 
The former case would explain the smallness of neutrino masses relative to those of the other fermions, owing to the suppression by the Majorana mass scale. This is the well-known and appealing case of the seesaw mechanism~\cite{Mohapatra:1979ia, Gell-Mann:1979vob, Yanagida:1979as, Minkowski:1977sc, Mohapatra:1980yp, Magg:1980ut, Lazarides:1980nt, Wetterich:1981bx, Foot:1988aq, Ma:1998dn}, which is also capable of explaining the observed baryon asymmetry in the universe~\cite{Yanagida:1979as}. The latter scenario, when the Majorana mass scale is suppressed relative to the electroweak scale, has received renewed attention recently~\cite{DeGouvea:2020ang,Martinez-Soler:2021unz,Sen:2022mun, Ansarifard:2022kvy, Chen:2022zts, Rink:2022nvw, Carloni:2022cqz,Franklin:2023diy, Perez-Gonzalez:2023llw, Dev:2024yrg}, due to its significantly different phenomenology. This alternative scenario, known as the pseudo-Dirac case, could arise from Planck-suppressed operators that violate total lepton number, and would lead to suppressed Majorana mass terms. From a phenomenological point of view, the pseudo-Dirac scenario gives rise to additional neutrino oscillations, not between flavour eigenstates, but rather between active and sterile states, leading to deficit of observable neutrinos depending on the mass splitting.

To put the possible effects of the MESM on firmer ground, we begin by rewriting the mass Lagrangian in~\eqref{eq:mass_lagr_1} as
\begin{equation}\label{eq:mass_lagr_2}
    \mathcal{L}_{\nu} = - \frac{1}{2} \overline{N_L^c} M N_L+{\rm h.c.}\, ,
\end{equation}
where
\begin{equation}
N_L = \begin{pmatrix}
\nu_L\\
(\nu_R)^c
\end{pmatrix},\quad
M =\begin{pmatrix}
0_3 & M_D^T\\
M_D & M_R^*
\end{pmatrix},
\end{equation}
with $M_D = v Y/\sqrt{2}$, $v$ the vacuum expectation value of the Higgs field, $\nu_L = (\nu_e, \nu_\mu, \nu_\tau)^T$ the vector of left-chiral neutrino fields, and $\nu_R=(\nu_{1,R}, \nu_{2,R}, \nu_{3,R})^T$ the vector of right-chiral neutrino fields. The diagonalisation of $M$ is performed in two steps. First, we define a $6\times 6$ matrix $V$ as
\begin{equation}
    V = \begin{pmatrix}
            V_L & 0 \\
            0 &  V_R^*
        \end{pmatrix},
\end{equation}
composed of the two unitary, $3\times 3$ matrices, $V_L$ and $V_R$, satisfying
\begin{equation}
    V_R^\dagger M_D V_L = M_D^{\rm diag},\qquad V_R^T M_R V_R = M_R^{\rm diag},
\end{equation}
with $M_D^{\rm diag} = {\rm diag}(m_{1,D}, m_{2,D}, m_{3,D})$ and $M_R^{\rm diag}={\rm diag}(m_{1,R}, m_{2,R}, m_{3,R})$. We assume that the same matrix, $V_R$, is responsible for the diagonalisation of both $M_D$ and $M_R$. This is justifiable, as we have the freedom to redefine the singlet states since they do not partake in any gauge interactions. Furthermore, this assumption implies that there will be only mixing between the $1-4$, $2-5$, and $3-6$ pairs. Applying the matrix $V$ to the mass matrix, we are left with
\begin{equation}
    V^T M V = \begin{pmatrix}
                0 & M_D^{\rm diag}\\
                M_D^{\rm diag} & M_R^{\rm diag}
              \end{pmatrix}=
              \begin{pmatrix}
                  0 & 0 & 0 & m_{1,D} & 0 & 0 \\
                  0 & 0 & 0 & 0 & m_{2,D} & 0 \\
                  0 & 0 & 0 & 0 & 0 & m_{3,D} \\
                  m_{1,D} & 0 & 0 & m_{1,R} & 0 & 0\\
                  0 & m_{2,D} & 0 & 0 & m_{2,R} & 0\\
                  0 & 0 & m_{3,D} & 0 & 0 & m_{3,R} 
              \end{pmatrix}
\end{equation}
where we have assumed that $M_R^{\rm diag}$ is real. This new matrix is still not diagonal, but it can be diagonalised in blocks, specifically the $1-4$, $2-5$, and $3-6$ blocks. This is achieved by applying a second matrix
\begin{equation}
    \vartheta =  \begin{pmatrix}
                      \enspace i C_\theta & S_\theta \\
                      -i S_\theta & C_\theta
                  \end{pmatrix},
\end{equation}
where the $3\times 3$ matrices, $C_\theta$ and $S_\theta$, are defined as
\begin{equation}
    C_\theta = {\rm diag}(\cos\theta_1 , \cos\theta_2 ,\cos\theta_3), \qquad 
    S_\theta = {\rm diag}(\sin\theta_1, \sin\theta_2, \sin\theta_3),
\end{equation}
with
\begin{equation}
    \tan 2\theta_{i} = \frac{2m_{i,D}}{m_{i,R}},
\end{equation}
defining the mixing angle, $\theta_i$, between the light and heavy neutrino mass eigenstates.
Thus, the full diagonalisation of $M$ is achieved through the matrix
\begin{equation}\label{eq:V_diag}
    \mathscr{V} = V \cdot \vartheta.
\end{equation}
The  {left-chiral components of the} mass eigenstates are given by
\begin{equation}
    N_L^{\rm diag} = (\nu_1^-, \nu_2^-, \nu_3^-, \nu_1^+, \nu_2^+, \nu_3^+)^T,
\end{equation}
satisfying $N_L = \mathscr{V}N_L^{\rm diag}$, where we use $\pm$ to label the mass of the neutrino eigenstate,
\begin{equation}\label{eq:masses_DR}
    m_{i}^\pm = \frac{1}{2}\left[\sqrt{(m_{i,R})^2+(2 m_{i,D})^2} \pm m_{i,R}\right].
\end{equation}
Finally, we note that we have fixed the CP phases in~\eqref{eq:V_diag} such that the eigenstates have positive masses.

We compute our neutrino mass spectrum as follows. Assuming normal ordering, with $m_{1}^{\pm} < m_{2}^\pm < m_{3}^\pm$, we first compute the mass of the lightest neutrino pair $m_1^\pm$ using~\eqref{eq:masses_DR}, with $m_D \equiv m_{1,D}$ and $m_R\equiv m_{1,R}$. Next, we compute the masses of the heavier neutrino pairs, $m_{2,3}^{\pm}$, { assuming the same squared mass splittings for both $m_i^-$ and $m_i^+$}, and using the observed solar and atmospheric mass splittings, $(m_{2}^\pm)^2 - (m_{1}^\pm)^2 = \Delta m_{\odot}^2 = 7.41\times 10^{-5}~{\rm eV^2}$, and $(m_{3}^\pm)^2 - (m_{1}^\pm)^2 = \Delta m_{\rm atm}^2= 2.51\times 10^{-3}~{\rm eV^2}$. This fixes the remaining values of $m_{i,D}$ and $m_{i,R}$. For the squared mass splittings, we take the latest values from~\cite{Esteban:2024eli}. {  We note that this assumption will have no significant impact on our conclusions. However, by reducing the number of free parameters from four to two, our assumption leads to more interpretable expressions and eliminates redundant sensitivity plots for each $m_{D,i}$ - $m_{R,i}$ pair, which would differ only by a scale factor and add no new insights.}

The neutrino fields in the flavour basis can be found by inspecting the charged current (CC) Lagrangian after rewriting the left-chiral neutrino fields in terms of the mass eigenstates
\begin{equation}\label{eq:CCL}
        \begin{split}
        \mathcal{L}_{\rm CC} &= -\frac{g}{\sqrt{2}}\sum_{\alpha,i}\overline{\ell_{\alpha L}}\gamma^\mu U_{\alpha i}(i \cos\theta_{i}\, \nu_i^- + \sin\theta_{i}\,\nu_i^+) W_\mu^\dagger + {\rm h.c.} \\
        &=-\frac{g}{\sqrt{2}}\sum_{\alpha} \overline{\ell_{\alpha L}}\,\gamma^\mu\,\nu_\alpha W_\mu^\dagger + {\rm h.c.},
        \end{split}
\end{equation}
where $g$ is the $SU(2)_L$ gauge coupling, and we immediately see that the flavour eigenstates are constructed from the mass eigenstates according to
\begin{equation}\label{eq:nu_active}
    \nu_{\alpha} = \sum_{i}U_{\alpha i} (i \cos\theta_{i}\nu_i^- + \sin\theta_{i} \nu_i^+).
\end{equation}
Here, we have defined $U_{\alpha i} = (V_L^\ell)^\dagger \cdot V_L$ as the standard Pontecorvo-Maki-Nakagawa-Sakata (PMNS) mixing matrix, with $V_L^\ell$ the unitary matrix that diagonalises the charged lepton sector. 
As demonstrated in~\cite{Perez-Gonzalez:2023llw}, the field operator $\bar{\nu}_i$ almost exclusively creates left-helicity neutrinos and annihilates right-helicity neutrinos in the ultrarelativistic limit, whilst its counterpart $\nu$ does the opposite.
As a result, the early universe will be populated by flavour eigenstates in the following combinations
\begin{align}\label{eq:states}
    |\nu_\alpha (h_P=-1)\rangle &= \sum_i U_{\alpha i}^*\, (-i \cos\theta_{i}|\nu_i^-\rangle + \sin\theta_{i}|\nu_i^+\rangle)\\
    |\nu_\alpha (h_P=+1)\rangle &= \sum_i U_{\alpha i}\, (i \cos\theta_{i}|\nu_i^-\rangle + \sin\theta_{i}|\nu_i^+\rangle),
\end{align}
where we use $h_P$ to denote the helicity of the eigenstate in the early universe. These linear combinations can also be thought of as neutrinos ($h_P = -1$) and antineutrinos ($h_P = +1$). This identification is particularly clear when comparing the Stodolsky effect in the MESM to the SM prediction for Dirac neutrinos, as we will see. In the absence of significant interactions, or if neutrinos are light enough so as not to cluster, the helicity profile of the C$\nu$B in the present day should match that of the early universe. As we will see, however, it will be important to distinguish the helicity at production from the helicity in the present day, as many effects depend differently on the two, and so we will leave the present day helicity as a free parameter.

So far we have not imposed any hierarchy between $M_D$ and $M_R$. Let us now consider how the two main limit scenarios affect the masses and linear combinations described above.
\begin{itemize}
    \item \emph{Pseudo-Dirac limit:} When $m_{i,R} \ll m_{i,D}$, neutrinos have masses
    \begin{align}
        m_i^\pm = m_{i,D} \pm \frac{m_{i,R}}{2},
    \end{align}  
    corresponding to a mass difference $\delta m_i \equiv m_i^+ - m_i^- = m_{i,R}$ between each pair. As this mass splitting is much smaller than the masses themselves, neutrinos will mostly act as Dirac fermions, with the possibility of active-sterile oscillations. The states from weak interactions will be maximally mixed, $\theta_i = \pi/4$, combinations of $|\nu_i^\pm\rangle$,
    \begin{subequations}
        \begin{align}
            |\nu_\alpha (h_P=-1)\rangle &= \frac{U_{\alpha i}^*}{\sqrt{2}}\, (-i |\nu_i^-\rangle + |\nu_i^+\rangle)\\
            |\nu_\alpha (h_P=+1)\rangle &= \frac{U_{\alpha i}}{\sqrt{2}}\, (i |\nu_i^-\rangle + |\nu_i^+\rangle).
        \end{align}
    \end{subequations}
    In the limit $m_{i,R}\to 0$, we identically recover Dirac neutrinos after recalling that a neutral Dirac fermion is a maximally mixed superposition of two identical mass Majorana fermions with opposite CP phases. 

    \item \emph{Seesaw}. If, on the other hand, $m_{i,R} \gg m_{i,D}$, the minus state masses are suppressed with respect of those of the plus states,
    \begin{align}
        m_i^-= \frac{(m_{i,D})^2}{m_{i,R}},\quad m_i^+ = m_{i,R},
    \end{align}    
    and the minus states will act as the mass eigenstates present in standard neutrino oscillations   
    \begin{subequations}
        \begin{align}\label{eq:stat}
            |\nu_\alpha (h_P=-1)\rangle &= -i  U_{\alpha i}^*\, |\nu_i^-\rangle,\\
            |\nu_\alpha (h_P=+1)\rangle &= i U_{\alpha i}\, |\nu_i^-\rangle.
        \end{align}
    \end{subequations}

\end{itemize}

A consequence of introducing three right-chiral singlets with a Majorana mass term is that we can no longer define the mass eigenstates in the same way as is done in standard neutrino oscillation treatments. However, we note that the quadratic mass splitting between the plus and minus pairs, $\delta m_i^2 \equiv (m_i^+)^2 - (m_i^-)^2 = 2m_{i,R}m_{i,D}$, must be either much larger or smaller than the experimentally measured mass differences between the three neutrino generations.
This is because the existence of additional mass differences in the interval $\delta m_i^2 = [10^{-12}, 10^2]~{\rm eV^2}$ is experimentally excluded, as it would modify the oscillation patterns that are well fitted by the $3-\nu$ framework\footnote{Even though some anomalies persist, such as the LSND, MiniBooNE, or BEST anomalies, it is not clear whether they originate from additional oscillations into sterile states, see e.g.~\cite{Hardin:2022muu}}.
As a result, we expect that standard oscillations between flavour eigenstates will occur on time scales different from those between the active and sterile states, especially in the pseudo-Dirac limit. 
 {
In particular, given the magnitude of the mass splittings between the three neutrino generations, it is well established that the flavour eigenstates decohered long before the present day~\cite{Long:2014zva}.  
Consequently, we define the following linear combinations, which we refer to as \textit{conventional} mass states,
\begin{equation}\label{eq:convential_states}
        |\nu_i (h_P)\rangle = i h_P \cos\theta_{i}|\nu_i^-\rangle + \sin\theta_{i}|\nu_i^+\rangle.
\end{equation}
These are related to the flavour eigenstates by $|\nu_\alpha (h_P=-1)\rangle = \sum_i U_{\alpha i}^*\,|\nu_i (h_P=-1)\rangle$, and $|\nu_\alpha (h_P=+1)\rangle = \sum_i U_{\alpha i}\, |\nu_i (h_P=+1)\rangle$. We stress, however, that these linear combinations do not possess definite masses unless $m_{R_i} = 0$, which corresponds to the Dirac limit.
These states serve as the initial conditions for the neutrino fluxes arriving at Earth\footnote{ {The combinations produced in CC interactions will always be those in~\eqref{eq:states}. As neutral current interactions are flavour diagonal, they will instead produce the conventional states defined in~\eqref{eq:convential_states}.
Owing to decoherence, this distinction makes no difference.}}.}

As a result of their different masses, the each of the six neutrino mass eigenstates states evolve with a different phase $\Phi_i^\pm(z)$, such that
\begin{align*}
    |\nu_i^\pm (z) \rangle= \exp\left(-i\Phi^\pm(z)\right) |\nu_i^\pm \rangle, 
\end{align*}
where $z$ defines the neutrino redshift at decoupling. Including the effects of the expansion of the universe~\cite{Weiler:1994hw, Beacom:2003eu,Esmaili:2012ac}, this phase is given by
\begin{align}\label{eq:phase}
    \Phi_i^{\pm}(z)=\int_0^{z} \frac{dz^\prime}{H(z^\prime)(1+z)} \left[(m_{i}^{\pm})^2 + p_{i}^2 (1+z^\prime)^2\right]^{\frac{1}{2}},
\end{align}
where $p_i$ is the present day momentum of relic neutrinos. The Hubble parameter, $H(z) = H_0\sqrt{\Omega_m (1+z)^3+\Omega_r(1+z)^4+\Omega_\Lambda}$, is given in terms of the Hubble constant, $H_0$, and the present day density parameters of non-relativistic matter, $\Omega_m$, radiation $\Omega_r$, and dark energy, $\Omega_\Lambda$. In what follows, we will use the parameters as determined by Planck~\cite{Planck:2018vyg,ParticleDataGroup:2020ssz}.

An accurate computation of the phase~\eqref{eq:phase} requires extreme numerical precision, as the difference in the two neutrino energies needs to be known to $\mathcal{O}(m_R)$. This is incredibly challenging, as during neutrino decoupling in the early universe, corresponding to the upper bound of the redshift integral, $p_{i} \sim \mathcal{O}(\mathrm{MeV})$. We therefore require relative precision of
\begin{equation}\label{eq:precision}
    \left(\frac{m_{R,\mathrm{min}}}{1\,\mathrm{MeV}}\right)^2 \simeq 10^{-80} \left(\frac{m_{R,\mathrm{min}}}{10^{-35}\,\mathrm{eV}}\right)^2,
\end{equation}
to accurately compute the neutrino phases over the full parameter space. For this reason, we use the \texttt{mpmath} package~\cite{mpmath} throughout, with decimal precision set by~\eqref{eq:precision}.

\begin{figure}[]
\centering
    \includegraphics[width = 0.75\linewidth]{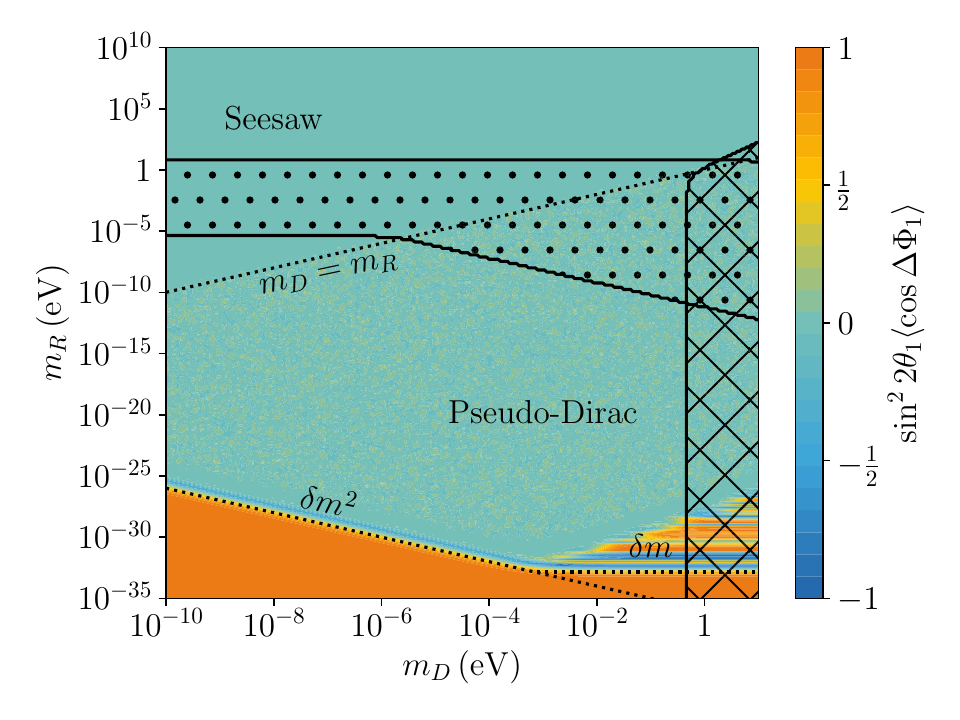}
    \caption{Averaged cosine of the phase difference $\Delta\Phi$ times $\sin^2 2\theta$ for the lightest neutrino pair, $\nu_1^\pm$, in the plane spanned by $m_R$ and $m_D$. We separate the limiting regions, \textit{i.e.} seesaw and pseudo-Dirac, with a dotted line. See the main text for a description of each region. We also present the experimentally excluded regions from oscillations (dotted region) and KATRIN (hatched region).}
    \label{fig:oscillation_contours}
\end{figure}
We show the cosine of the phase difference, $\Delta\Phi_i \equiv \Phi_i^+-\Phi_i^-$, weighted by $\sin^2 2 \theta_i$, over the plane spanned by $m_D$ and $m_R$ for the lightest neutrino generation\footnote{We plot our results in the normal mass hierarchy throughout, with $m_1^\pm < m_2^\pm < m_3^\pm$.} in Figure~\ref{fig:oscillation_contours}. This is averaged over a normalised Fermi-Dirac distribution at the present day C$\nu$B temperature, $T_{\nu,0} = 0.168~{\rm meV}$, 
\begin{equation}
    \bar{f} (p_{i}) = \frac{2}{3\zeta(3) T_{\nu,0}^3} \frac{1}{\exp\left(\frac{p_{i}}{T_{\nu,0}}\right)+1},
\end{equation}
explicitly
\begin{align}
    \sin^2 2 \theta_i\,\langle \cos\Delta\Phi_i \rangle = \sin^2 2 \theta_i\,\int_0^\infty\, dp_i\, p_i^2\, \bar{f} (p_{i})\, \cos\Delta\Phi_i(p_i).
\end{align}
We have chosen to present such a specific combination as it is the one that will appear in the detection observables, and its absolute value is effectively a measure of how closely the neutrinos resemble a Dirac fermion for each combination of $m_D$ and $m_R$.
In Figure~\ref{fig:oscillation_contours} there is a clear distinction between the pseudo-Dirac and seesaw regions. These regions are bounded by the line $m_R=m_D$, and the distinction is predominantly due to the term $\sin^2 2 \theta$, which tends to one in the Dirac limit. Conversely, when the mixing between the plus and minus states pair is negligible, we recover the seesaw limit.
In Figure~\ref{fig:oscillation_contours}, we also highlight the regions already excluded by experiments. The region marked with dots indicates the parameters that would induce active-sterile oscillations that could be observed in flavour oscillation experiments, whilst the hatched region indicates the values of active neutrino masses inconsistent with the recent KATRIN results~\cite{Katrin:2024tvg}.
In the region where $m_R \lesssim m_D$, and above the dotted curves labelled by $\delta m^2$ and $\delta m$, the mixing angle is maximal, but the phase difference differs significantly for each momentum value, such that the average over the Fermi-Dirac distribution is approximately zero. We find a more interesting behaviour when $m_R\ll m_D$, that is, in the extreme pseudo-Dirac limit. 
To understand this limit, let us note that the expression for the phase~\eqref{eq:phase} has no general expansion in terms of small parameters. 
However, when the neutrinos are either ultra- or non-relativistic today, we can make the expansions\footnote{In the non-relativistic limit, the expansion $m_{i}^\pm \gg \left\langle p_{i} \right\rangle (1+z)$ breaks down in the early universe as $z \to z_\mathrm{dec}$. However, as the integral is heavily dominated by the regions with $z \lesssim 1000$, where the expansion holds, we can still safely use the expansion.}
\begin{equation}\label{eq:phase_expansions}
    \Delta \Phi_i = \int_{0}^{z_\mathrm{dec}}\frac{dz}{(1+z)H(z)}\Delta E_i(z) \simeq \begin{cases}
        \frac{\delta m^2}{2\left\langle p_{i}\right\rangle} L_2 , &\qquad  \text{ultra-relativistic}, \\
        \delta m \,L_1, &\qquad \text{non-relativistic},
    \end{cases}
\end{equation}
where $1 + z_\mathrm{dec} = T_{\nu,\mathrm{dec}}/T_{\nu,0} \simeq 5.95\cdot 10^{9}$ defines the redshift at neutrino decoupling, assuming neutrino decoupling at a temperature $T_{\nu,\mathrm{dec}} = 1\,\mathrm{MeV}$, and $\left\langle p_{i}\right\rangle = 3.15\,T_{\nu,0} \simeq 0.529\,\mathrm{meV}$ is the mean momentum of relic neutrinos today. In~\eqref{eq:phase_expansions}, we have defined the distances~\cite{DeGouvea:2020ang} 
\begin{equation}\label{eq:distanceMeasures}
    L_n = \int_0^{z_\mathrm{dec}}\frac{dz}{(1+z)^n H(z)},
\end{equation}
such that $L_0$ is the comoving distance, whilst $n = 1$ corresponds to the distance travelled by a massless particle from redshift $z_\mathrm{dec}$ to the present day~\cite{Weiler:1994hw, Beacom:2003eu,Esmaili:2012ac}.
For small enough mass splittings, we therefore expect the phase difference $\Delta \Phi_i \to 0$. In the ultra-relativistic case, this occurs when
\begin{equation}
    \delta m^2 \ll \frac{2\left\langle p_{i}\right\rangle}{L_2} \simeq 2.89 \cdot 10^{-36}\,\mathrm{eV}^2,
\end{equation}
or equivalently when the product of the Dirac and Majorana masses lies below the hyperbola defined by
\begin{equation}
    m_D m_R = 1.44 \cdot 10^{-36}\,\mathrm{eV}^2.
\end{equation}
In the non-relativistic case, the phase difference is instead minimised for
\begin{equation}
    \delta m \simeq m_R \ll \frac{1}{L_1} \simeq 1.51\cdot 10^{-33}\,\mathrm{eV},
\end{equation}
\textit{i.e.} when the mass splitting is approximately equal to the inverse of the distance travelled by a massless particle since neutrino decoupling. In this region, neutrinos will behave exactly as Dirac fermions, as all lepton number violating processes proceed at vanishing rates, and relic neutrino states, the neutrinos with longest oscillation baseline of all, will not have deviated significantly from their early universe superpositions. The C$\nu$B is therefore the ultimate probe of the Dirac or Majorana nature of neutrinos, capable of setting the strongest constraints on the Majorana mass, $m_R \lesssim 10^{-33}\,\mathrm{eV}$.

%
\section{C$\nu$B detectors}\label{sec:detectors}

Each of the many proposals to search for relic neutrinos has a unique dependence on the neutrino mass, giving rise to a rich landscape of detection sensitivities~\cite{Bauer:2022lri}.  In this section we will derive several of these sensitivities, explicitly those of PTOLEMY, the Stodolsky effect, coherent scattering, and an accelerator experiment in the MESM, which gives rise to additional mass and helicity dependencies. 

There are two key differences between the relic neutrino detection sensitivities in the SM and the MESM:
first, the incoming neutrino flux will consist of a mixture of the true mass eigenstates, $\nu_{i}^\pm$, which may differ from the initial superpositions $|\nu_i (h_P)\rangle$ as defined in~\eqref{eq:convential_states}. Second, neither neutral current (NC) nor charged current (CC) interactions are diagonal in the $\nu_i^\pm$ basis, potentially leading to transitions between these states, analogous to flavour-changing processes in non-standard interaction (NSI) scenarios.

To account for each of these effects, we adopt a density matrix formalism similar to that used in~\cite{Coloma:2022umy} and~\cite{Amaral:2023tbs} for the case of NSI. 
This framework accurately captures the interplay between propagation and scattering in a non-diagonal basis and leads to correct predictions for the observables, as we will demonstrate.
The general expression for a relic neutrino observable ${\cal O}$ in the MESM is
\begin{equation}\label{eq:full_f}
    \mathcal{O} = \sum_\mathrm{d.o.f.} \mathcal{C}\, n_{\nu}(\nu_{i,h})\, \Tr [K \rho_i(h_P,z)].
\end{equation}
Here $\mathcal{C}$ is some non-kinematic prefactor, which may also depend on the neutrino degrees of freedom, such as the present day helicity, $h$, or neutrino generation, $i \in \{1,2,3\}$, and where
\begin{equation}
    n_{\nu}(\nu_{i,h}) = \frac{3 \zeta (3)}{4 \pi ^2} T_{\nu,0}^3 \simeq 56\,\mathrm{cm}^{-3},
\end{equation}
is the number density per degree of freedom\footnote{The number densities of the two helicity eigenstates are only equal if neutrinos are Majorana fermions. See~\cite{Long:2014zva} and~\cite{Bauer:2022lri} for a comprehensive discussion.}, given in terms of the present day relic neutrino temperature $T_{\nu,0} = 0.168\,\mathrm{meV}$. The matrix $K$ is the expectation value of some process-dependent kinematic operator, which we will define on a process-by-process basis, and $\rho_i(h_P,z)$ is the density matrix for the incoming C$\nu$B flux emitted at a redshift $z$ with helicity $h_P$, which accounts for the evolution of the neutrino states as they propagate. The elements of the density matrix in the $|\nu_i^\pm\rangle$ basis are given by 
\begin{equation}
    \begin{split}
        \rho_i^{xy}(h_P,z) &= \braket{\nu_i^x(h_P,0)|\nu_i(h_P,z)} \braket{\nu_i(h_P,z)|\nu_i^y(h_P,0)},
    \end{split}
\end{equation}
witn $x,y \in \{+,-\}$, such that full matrix takes the form
\begin{equation}
    \rho_i(h_P,z) = \begin{pmatrix}
        \rho_i^{--} & \rho_i^{-+}  \\
        \rho_i^{+-} & \rho_i^{++}
    \end{pmatrix} = \begin{pmatrix}
                        \cos^2\theta_i & i h_P \cos\theta_i\sin\theta_i\, e^{i\Delta \Phi_i} \\
                         -i h_P \cos\theta_i\sin\theta_i\, e^{-i\Delta \Phi_i} &\sin^2\theta_i
                  \end{pmatrix}. 
\end{equation}
After performing the trace in~\eqref{eq:full_f}, we therefore find the most general observable in MESM
\begin{equation}\label{eq:masterEq}
    \begin{split}
    \mathcal{O} &= \sum_{\mathrm{d.o.f.}} \mathcal{C} \,n_{\nu}(\nu_{i,h}) \left[\sin^2\theta_i K^{++} +  \frac{ih_P}{2}\sin 2\theta_i\left(K^{+-}e^{i\Delta\Phi} - K^{-+}e^{-i\Delta\Phi}\right)+ \cos^2\theta_i K^{--}\right] \\
    &=\sum_{\mathrm{d.o.f.}} \mathcal{C} \,n_{\nu}(\nu_{i,h}) \left[\sin^2\theta_i K^{++} -h_P \sin 2\theta_i\,\mathrm{Im}\left(K^{+-}e^{i\Delta\Phi}\right)+ \cos^2\theta_i K^{--}\right], 
    \end{split}
\end{equation}
where we have used $K^{-+} = (K^{+-})^*$ in going from the first to the second line, which follows from the hermiticity of $\hat{K}$, required for $\mathcal{O}$ to be an observable.
This immediately recovers the result that only the light neutrino states contribute in the seesaw limit, $\theta_i \to 0$, where the heavy states coincide almost exactly with the sterile singlet states. In the off-diagonal elements of $K$, we will encounter terms of the form $u(p_{i}^+, h) \bar{u}(p_{i}^-, h)$, with $u$ the positive frequency Dirac spinor, and analogous combinations of its negative helicity counterpart, $v$. In general, computing the value of such terms is difficult. To simplify things, we can leverage the fact that these terms are always multiplied by $\sin\theta_i$, and will vanish in the seesaw limit. As a result, we only need to worry about them in the pseudo-Dirac regime, where $m_i^+ \simeq m_i^-$ and we can use the identities with an effective neutrino mass $\bar{m}_i \equiv (m_i^+ + m_i^-)/2$. We will denote such quantities with bars throughout. In the intermediate region, where $m_{i,D} \simeq m_{i,R}$, this formalism becomes less accurate, however, as this is both a small region, and much of it is excluded by oscillation experiments, this is of less concern. When plotting the sensitivity of each of these proposals, we will therefore shade these regions to highlight where our formalism may be less accurate.  {Specifically, we will highlight the regions where 
\begin{equation}\label{eq:uncertaintyApprox}
    \sin^2 2\theta_i \left(\frac{m_i^+ - m_i^-}{m_i^+ + m_i^-}\right) \geq 0.1.
\end{equation}
The quantity on the LHS of~\eqref{eq:uncertaintyApprox} is an estimate of the leading order uncertainty introduced by making the averaged mass approximation for the off-diagonal terms.
}

To determine the matrix elements of the operator $\hat{K}$ in the $|\nu_i^\pm\rangle$ basis, we classify the detection processes into two main categories. 
The first category includes processes related to neutrino scattering, such as neutrino capture, explicitly PTOLEMY, coherent scattering, and accelerator experiments, where the associated operator is proportional to the cross section for the process. 
The second category involves modifications to the energy levels of atomic electron spin states caused by the presence of a neutrino background, known as the Stodolsky effect. 
In this case, the operator $\hat{K}$ is the expectation value of the effective low-energy neutrino-electron Hamiltonian. 
Let us examine these two categories in more detail.
\begin{enumerate}

    \item \textbf{Scattering processes:} In this case, we again follow~\cite{Coloma:2022umy} and~\cite{Amaral:2023tbs}, and define  $\hat{K}\propto\hat{\sigma}$, with $\hat\sigma$ the generalised cross section relevant for the process, having matrix elements $\sigma^{xy}$ with $x,y \in \{+,-\}$, in the $|\nu_i^\pm\rangle$ basis. For a $\nu_i^x + P \to f + D$ process, in which $f$ could be either a neutrino or charged lepton and $P,D$ are parent and daughter nuclei, such matrix elements are obtained via
    \begin{align}
        \frac{d\sigma^{xy}}{dt}~\propto~ {\cal M}^*(\nu_i^x + P \to f + D) {\cal M}(\nu_i^y + P \to f + D),
    \end{align}
    where $t$ is the usual Mandelstam variable, and we have omitted the phase-space factors for brevity. We additionally note that when the final state is a neutrino, \textit{i.e.} $f = \nu_i^z$, a sum over $z \in \{+,-\}$ should be performed. To construct each of these amplitudes, and importantly get the correct sign of $i\cos\theta_i$, we follow the rules set out in Appendix~\ref{app:mixingFactors}.

    \item\textbf{Stodolsky effect:} Since the Stodolsky effect is proportional the expectation value of the neutrino-electron Hamiltonian ${H}_{\rm int}$ in a relic neutrino background, we will simply have that $K = {H}_{\rm int}$. The matrix elements of ${H}_{\rm int}$ will be
    \begin{align}
        {H}_{xy}(s_e) = \langle e_{s_e}, \nu_{i}^y |\, {H}_{\rm int}\, | e_{s_e}, \nu_{i}^x \rangle\, ,
    \end{align}
    with $| e_{s_e}\rangle$ the electron states with spin $s_e$. 
    
\end{enumerate}
Before examining each detection proposal, we make one final, but important comment about the neutrino helicity. In the absence of significant interactions between decoupling, and provided that neutrinos are not massive enough to cluster\footnote{The minimum mass for neutrinos to cluster has been estimated in~\cite{Shergold:2021evs} as $m_i \gtrsim 0.29\,\mathrm{eV}$. As neutrinos of this mass are disfavoured by cosmological bounds~\cite{DESI:2024mwx}, we will assume that neutrinos do not cluster.}, their helicity in the present day will be the same as that in the early universe. However, we note that helicity is not a Lorentz invariant parameter, and that transformations between the C$\nu$B reference frame and the laboratory frame, estimated to be moving at a speed $\beta_\Earth \simeq 10^{-3}$ relative to the C$\nu$B frame~\cite{Bauer:2022lri}, can change the helicity. As such, we will only use $h_P$ to denote the linear superpositions of $\ket{\nu_i^\pm}$ that appear in the density matrix, and $h$ when dealing with observables in $K$. Consequently, we will need to know the relations between the neutrino fluxes in the C$\nu$B frame and the lab frame. Following~\cite{Bauer:2022lri}, the helicity dependent fluxes are related in the two frames by
\begin{equation}\label{eq:helicityFlip}
    n_\nu(\nu_{i,h}) = \delta_{h,h_P} (1-\mathcal{P}_i) n_\nu(\nu_{i,h_P}) + (1-\delta_{h,h_P}) \mathcal{P}_i n_\nu(\nu_{i,h_P'}),
\end{equation}
where $h_P' \neq h $, and $\mathcal{P}_i$ is the probability of a neutrino flipping helicity given by
\begin{equation}
    \mathcal{P}_i = \frac{1}{2}\left[1 - \mathcal{H}(\beta_i-\beta_\Earth)\left(1 - \frac{2}{\pi}\arcsin\left(\frac{\beta_\Earth}{\beta_i}\right)\right)\right],
\end{equation}
where $\mathcal{H}$ denotes the Heaviside step function. Any helicity asymmetry present in the C$\nu$B in its own reference frame will therefore completely vanish for those momenta corresponding to $\beta_i \leq \beta_\Earth$, corresponding to $\mathcal{P}_i \to \frac{1}{2}$. As we will see, this will have particularly important consequences for the Stodolsky effect. Finally we note that when transforming $n_\nu$ between frames, we will use the velocity corresponding to the neutrino mass corresponding to the type of term that they multiply, $m_i^+$ if they multiply $\rho^{++}_i$, $m_i^{-}$ if they multiply $\rho_i^{--}$, and $\bar{m}_i$ if they multiply $\rho^{\pm \mp}_i$. We again justify this choice by noting that the $\rho^{\pm \mp}_i$ terms will vanish in the seesaw limit, and so the choice is irrelevant there, and that in the pseudo-Dirac limit, the masses are almost degenerate. 

We will now examine each detection proposal, highlighting its specific features within the framework of the density-matrix approach.

\subsection{PTOLEMY}
\begin{figure}
    \centering
    \includegraphics[width=\textwidth]{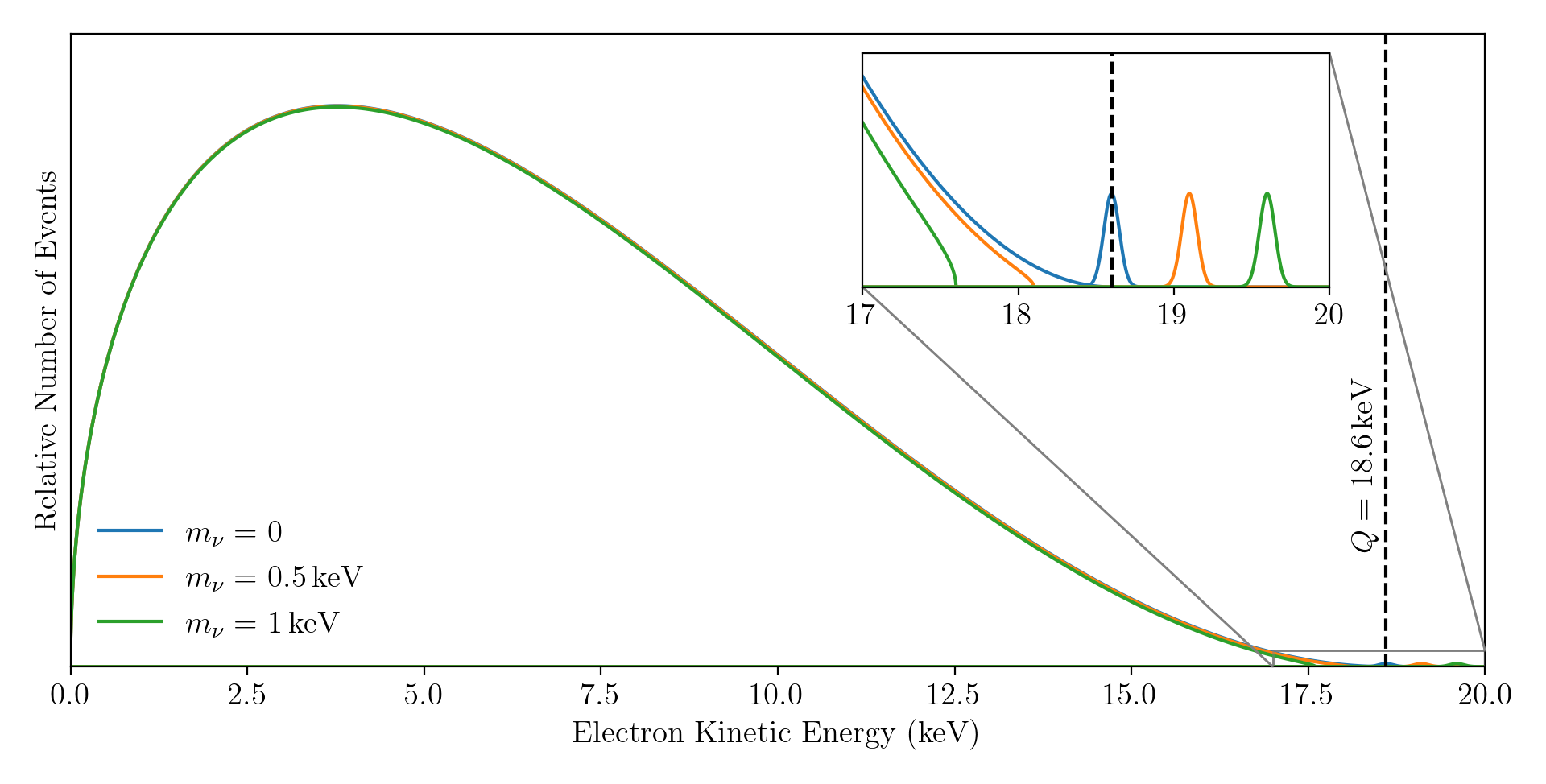}
    \caption{Toy electron energy spectrum at a PTOLEMY-like experiment, for the exaggerated neutrino masses $m_{\nu} = 0\,\mathrm{keV}$ (blue), $0.5\,\mathrm{keV}$ (orange), and $1\,\mathrm{keV}$ (red). The electrons with kinetic energy less than the beta decay energy, $Q = 18.6\,\mathrm{keV}$, originate from tritium beta decays, whilst those to the right are due to relic neutrino capture.}
    \label{fig:ptolemySpectrum}
\end{figure}

We first turn our attention to the PTOLEMY experiment~\cite{Long:2014zva,Bauer:2022lri,Weinberg:1962zza,Cocco:2007za,Betts:2013uya,Akhmedov:2019oxm,PTOLEMY:2022ldz}, where the idea is to capture an electron neutrino from the C$\nu$B on a tritium atom in the process
\begin{equation}
    \nu_e + {^3 \mathrm{H}} \to e^- + {^3 \mathrm{He}^+},
\end{equation}
and measure the energy of the outgoing electron with extreme precision. Upon absorbing a relic neutrino of energy $E_{i}$, the electron will be emitted with energy\footnote{Here we have neglected nuclear recoil energy for clarity. As this is expected to exceed the neutrino mass, it should be included when determining the experimental resolution at PTOLEMY. See~\cite{Long:2014zva} for more details.} $E_{\mathrm{C}\nu\mathrm{B}} \simeq Q + E_{i}$, with $Q \simeq 18.6\,\mathrm{keV}$ the energy released in tritium beta decay. After counting many electrons, including those from the natural beta decays of tritium, emitted with kinetic energy up to $E_\mathrm{max} \simeq Q - m_{i}$, we arrive at a spectrum analogous to the toy spectrum shown in Figure~\ref{fig:ptolemySpectrum}. The appearance of peaks displaced from the beta decay endpoint energy by $\sim 2 m_{i}$ is therefore a clear indicator of relic neutrino capture.

To determine the expected neutrino capture rate at PTOLEMY in the MESM, we compute
\begin{equation}\label{eq:ptolemyRateTrace}
    \Gamma = N_T \sum_{i,h} n_\nu(\nu_{i,h})\,\Tr [\sigma_{\rm P}\,\rho_i(h_P,z)],
\end{equation}
where $N_T \simeq 2 \cdot 10^{25}$ the number of atoms in $m_T = 100\,\mathrm{g}$ of tritium, $h = 1$ for right-helicity neutrinos, $h = -1$ for left-helicity neutrinos, and where $\sigma_{\rm P}$ is the generalised cross section for PTOLEMY. The generalised cross section for neutrino capture on tritium is
\begin{equation}
    \sigma_\mathrm{P} = \begin{pmatrix}
        \sigma_\mathrm{P}^{--} & \sigma_\mathrm{P}^{-+} \\
        \sigma_\mathrm{P}^{+-} & \sigma_\mathrm{P}^{++}
    \end{pmatrix} = |U_{ei}|^2 \sigma_0 \begin{pmatrix}
        \cos^2\theta_i \,\mathcal{A}_{i,h}^{-} & -\frac{i}{2}\sin 2\theta_i \,\bar{\mathcal{A}}_{i,h} \\
        \frac{i}{2}\sin 2\theta_i \,\bar{\mathcal{A}}_{i,h} & \sin^2\theta_i\, \mathcal{A}_{i,h}^{+}
    \end{pmatrix},
\end{equation}
where $\sigma_0$ is the capture cross section, which for neutrinos with energies much less than a few $\mathrm{keV}$ takes the approximately constant value $\sigma_0 \simeq 3.84 \cdot 10^{-45}\,\mathrm{cm}^{2}$. The helicity dependent factor is
\begin{equation}
    \mathcal{A}_{i,h}^x = 1 - h \beta_i^x,
\end{equation}
with $\beta_i^x$ the neutrino velocity. Inserting the appropriate elements into~\eqref{eq:ptolemyRateTrace}, and using~\eqref{eq:masterEq}, we therefore find the capture rate
\begin{equation}\label{eq:ptolemyRate}
    \begin{split}
    \Gamma &= N_T \sum_{i,h,h_P}|U_{ei}|^2 \sigma_0\, n_\nu(\nu_{i,h}) \left[\cos^4\theta_i \mathcal{A}^-_{i,h} - \frac{h_P}{2}\sin^2 2\theta_i \cos\Delta \Phi_i \bar{\mathcal{A}}_{i,h} + \sin^4\theta_i \mathcal{A}_{i,h}^+ \right] \\
    &= 4.07\,\mathrm{y}^{-1} \left(\frac{m_T}{100~{\rm g}}\right) \sum_{i,h,h_P}|U_{ei}|^2 \, \left(\frac{n_\nu(\nu_{i,h})}{56\,\mathrm{cm}^{-3}}\right)\left[\cos^4\theta_i \mathcal{A}^-_{i,h} - \frac{h_P}{2}\sin^2 2\theta_i \cos\Delta \Phi_i \bar{\mathcal{A}}_{i,h} \right.\\
    &\qquad\qquad\qquad\qquad\qquad\qquad\qquad\qquad\qquad\qquad~\left. + \sin^4\theta_i \mathcal{A}_{i,h}^+ \right].
    \end{split}
\end{equation}
As all but the lightest generation must be non-relativistic if we assume a constant temperature $T_{\nu,0}$ for all three generations, we expect that $\mathcal{A}_{i,h} \sim 1$ in most scenarios, such that the neutrino capture rate is largely independent of the neutrino kinematics.

Now we can examine the two limits. In the seesaw limit, only the light neutrino terms survive, and we are left with
\begin{equation}
    \Gamma_{\mathrm{SS}} = N_T \sum_{i,h,h_P}|U_{ei}|^2 \sigma_0\, n_\nu(\nu_{i,h}) \mathcal{A}^-_{i,h},
\end{equation}
recovering the result for Majorana neutrinos given in~\cite{Long:2014zva} and~\cite{Bauer:2022lri}. On the other hand, in the pseudo-Dirac limit, we set $\mathcal{A}_{i,h}^+ \simeq A_{i,h}^- \simeq \bar{\mathcal{A}}_{i,h}$. This leaves us with
\begin{equation}
    \Gamma_\mathrm{PD} = N_T \sum_{i,h,h_P}|U_{ei}|^2 \sigma_0\, n_\nu(\nu_{i,h}) \bar{\mathcal{A}}_{i,h} \left[1 - \frac{1}{2}\sin^2 2\theta_i(1+h_P\cos\Delta\Phi_i)\right].
\end{equation}
After explicitly substituting in the production helicity, $h_P$, we see that the term in square brackets is precisely the probability of oscillating to a detectable state as given in~\cite{Perez-Gonzalez:2023llw}
\begin{equation}\label{eq:probs_ptol}
    \begin{split}
    P(\nu_{i,h_P} \to \nu_{i,d}) &= 1 - \frac{1}{2}\sin^2 2\theta_i(1+h_P\cos\Delta\Phi_i) \\
    &= 1 - \sin^2\,2\theta_{i} \times
    \begin{cases}
        \sin^2\left[\frac{\Delta \Phi_i}{2}\right], &\qquad h_P=-1,\\
        \cos^2\left[\frac{\Delta \Phi_i}{2}\right], &\qquad h_P=+1.
    \end{cases}
    \end{split}
\end{equation}
The appearance of the oscillation probability indicates that the simplified method of calculating the number of events using a probability-weighted cross section holds when the detection process is diagonal in the $\pm$ basis. In the case of neutrino capture, where the final state is a charged lepton, there are no transitions between $\pm$ states and the standard approach remains valid. The specific dependence on production helicity in the oscillation probability~\eqref{eq:probs_ptol} arises because the relevant term in the CC Lagrangian for the capture process is proportional to $\overline{e_L} \gamma^\mu \nu_i^\pm$. In the Dirac limit, this term annihilates neutrinos but not antineutrinos. This is reflected in the probability by setting $\Delta \Phi_i \to 0$ and $\theta_i \to \pi/4$, leading to $P(\nu_{i,R} \to \nu_{i,d}) = 0$.
Although seemingly counterintuitive, this zero probability indicates that right-helicity neutrinos remain undetectable for PTOLEMY in the extreme pseudo-Dirac limit. This, in turn, recovers the capture rate for Dirac neutrinos given in~\cite{Long:2014zva} and~\cite{Bauer:2022lri}. Likewise, if the CC interaction process is sensitive solely to antineutrinos and not neutrinos, the probability $P(\nu_{i,L} \to \nu_{i,d})$ will approach zero in the Dirac limit. This situation will become relevant in the case of the accelerator proposal for C$\nu$B detection, as we shall discuss.

Despite the promising $\sim 4$ events expected per year at PTOLEMY, there are many as-yet-unresolved issues to overcome before a successful experiment can be performed. This is due a combination of two factors: extreme energy resolution $\Delta \lesssim 2 m_{i}$ is required to distinguish the electrons due to neutrino capture from tritium beta decay electrons; the incredible challenge of building a detector with a $100\,\mathrm{g}$ fiducial mass of tritium. Individually, solutions to these challenges have been proposed by the PTOLEMY collaboration, who project a sensitivity of $\Delta \simeq 0.05\,\mathrm{eV}$ using the drift filter presented in~\cite{Betti:2018bjv,Apponi:2021hdu}, and that the required $100\,\mathrm{g}$ mass of tritium can be stored and kept active by binding it to a graphene substrate~\cite{PTOLEMY:2019hkd,PTOLEMY:2022ldz}. Unfortunately, as first discussed in~\cite{Cheipesh:2021fmg} and later expanded upon in~\cite{Nussinov:2021zrj}, confining the tritium atoms to a small volume introduces an unavoidable uncertainty in the final state electron energy, which is larger than the proposed energy resolution of the detector. This, in turn, could diminish the efficacy of the PTOLEMY proposal by many orders of magnitude, either by requiring a smaller fiducial mass of tritium, lowering the expected event rate, or worsening the energy resolution if the original $100\,\mathrm{g}$ of tritium is used without overcoming the broadening issue, thus removing the ability to detect very light neutrinos. This issue has been addressed by the PTOLEMY collaboration in~\cite{PTOLEMY:2022ldz}, and several methods of reducing the induced uncertainty have been proposed. Nevertheless, as no definite solution has been presented, other proposals for relic neutrino detection should be explored.

\subsection{Stodolsky effect}\label{sec:stodolsky}
We now turn our attention to the Stodolsky effect~\cite{Stodolsky:1974aq,Duda:2001hd,Gelmini:2004hg,Domcke:2017aqj,Bauer:2022lri,Rostagni:2023eic}, where a SM fermion with couplings to neutrinos in a background of relic neutrinos experiences a spin-dependent energy shift, analogous to the Zeeman effect for a charged particle in a magnetic field. Where these effects differ, however, is that an asymmetry in the C$\nu$B is required in order for a non-zero energy shift due to the Stodolsky effect; no such asymmetry in the photon background is required for the Zeeman effect. 
This asymmetry can either be between the number of neutrinos and antineutrinos in the background, or between the number of present day left- and right-helicity neutrinos.

The magnitude of the energy splitting of the two electron spin states in a background of relic neutrinos due to the Stodolsky effect is obtained by taking the difference between the expectation values of the neutrino-electron interaction Hamiltonian~\cite{Bauer:2022lri, Rostagni:2023eic} for the two electron spins. The energy shift for each electron spin state is given by
\begin{equation}\label{eq:stodl_gen}
    \delta E(s_e) = \frac{1}{4m_e} \sum_{i,h,h_P} n_\nu(\nu_{i,h}) \left\langle\frac{1}{E_i^{-+}}\Tr[{H}_{\rm int}(s_e)\, \rho_i(h_P,z)]\right\rangle,
\end{equation}
where $E_i^{-+}\equiv\cos^2\theta_i E_i^-+\sin^2\theta_i E_i^+$, $m_e$ is the electron mass, and the angled brackets indicate averaging due to the relative motion of the Earth with respect to the C$\nu$B frame.
The matrix elements of the interaction Hamiltonian are
\begin{equation}
    {H}_{\rm int}(s_e) = 4\sqrt{2}G_F m_e A_{ii} s_e \begin{pmatrix}
        h\cos^2\theta_i\, m_i^-\,(S_e \cdot S_i^-)   & \frac{i}{2}\sin 2\theta_i \,(S_e\cdot \bar{p}_{i}) \\
        -\frac{i}{2}\sin 2\theta_i \,(S_e\cdot \bar{p}_{i}) & h\sin^2\theta_i\, m_i^+\,(S_e \cdot S_i^+)
    \end{pmatrix} + f(V_{ii}) \,,
\end{equation}
where $A_{ii} = g_{A,e} + |U_{ei}|^2$ is given in terms of the electron axial coupling to $Z$-bosons, $g_{A,e} = -\frac{1}{2}$, and $f(V_{ii})$ contains elements proportional the the vector coupling $V_{ii}$, which will cancel when taking the difference between the energy shifts. We will therefore neglect these terms in what follows. The electron and neutrino polarisation vectors, $S_e$ and $S_i^\pm$, respectively, are given by
{ \begin{equation}
    S_e^\mu = \left(0, \hat{n}\right), \qquad (S^\pm_i)^\mu = \left(\frac{|\vec p_i|}{m_i^\pm}, \frac{E_i^\pm}{m_i}\frac{\vec p_i}{|\vec p_i|}\right),
\end{equation}
where $\hat{n}$ is the unit vector along the electron spin axis, and we have assumed that the electron is at rest in the lab frame.}
Finally, the trace of the product of the density matrix and interaction Hamiltonian is given by,
\begin{equation}
    \begin{split}
    \Tr[{H}_{\rm int}(s_e)\, \rho_i(h,z)] = 4\sqrt{2}G_F A_{ii} &m_e s_e \, \bigg[ \frac{h_P}{2}\sin^2 2\theta_i \cos\Delta\Phi_i \,(S_e\cdot p_{i}) \\
    &+h\left(\cos^4\theta_i\, m_i^-\,(S_e \cdot S_i^-) + \sin^4\theta_i\, m_i^+\,(S_e \cdot S_i^+)\right)
    \bigg].
    \end{split}
\end{equation}
When performing the averaging in~\eqref{eq:stodl_gen}, we can simplify things significantly by once again considering the limits. In the seesaw regime, only the light neutrino terms will enter, so we can safely replace the denominators with $E^{-}_i$ for the terms proportional to $\mathcal{H}_\mathrm{int}^{--}$. This approximation still holds well in the pseudo-Dirac regime, where $E^+_i \simeq E^{-}_i$. We can play a similar trick for the remaining terms, those proportional to $\mathcal{H}_\mathrm{int}^{++}$ and $\mathcal{H}_\mathrm{int}^{+-}$, and set their denominators to $E^{+}_i$ and $\bar{E}_i$, respectively. We justify this by noting that since these terms will vanish in the seesaw limit, the choice of denominator is irrelevant there, and that in the pseudo-Dirac regime, we once again have $E^+_i \simeq E^{-}_i \simeq \bar{E}_i$. These approximations allow us use the results from~\cite{Bauer:2022lri}, yielding
\begin{equation}
    \begin{split}
    \left\langle \frac{1}{E_i^{-+}}\Tr[{H}_{\rm int}(s_e)\, \rho_i(h_P,z)]\right\rangle &\simeq -\frac{4\sqrt{2}G_F A_{ii}}{3} \beta_\Earth m_e s_e \, \bigg[h_P \sin^2 2\theta_i \cos\Delta\Phi_i \, \left(2-\bar{\beta}_i^2\right)  \\
    & \quad+ h \left(\frac{\cos^4\theta_i }{\beta_{i}^-}\left(3-(\beta_{i}^-)^2\right) + \frac{\sin^4\theta_i }{\beta_{i}^+}\left(3-(\beta_{i}^+)^2\right)\right) \bigg].
    \end{split}
\end{equation}
Inserting this into~\eqref{eq:stodl_gen}, and taking the difference between the two electron spin state gives us the magnitude of the energy splitting
\begin{equation}\label{eq:stodSplitting}
    \begin{split}
    \Delta E &\simeq \frac{2\sqrt{2}G_F}{3} \beta_\Earth \sum_{i,h,h_P} n_{\nu}(\nu_{i,h}) A_{ii}\, \bigg[h_P \sin^2 2\theta_i \cos\Delta\Phi_i \, \left(2-\bar{\beta}_i^2\right)  \\
    &\qquad\qquad\qquad\qquad\qquad\qquad+ h\left(\frac{\cos^4\theta_i }{\beta_{i}^-}\left(3-(\beta_{i}^-)^2\right) +\frac{\sin^4\theta_i }{\beta_{i}^+}\left(3-(\beta_{i}^+)^2\right)\right) \bigg] \\
    &= 4.73\cdot 10^{-39}\,\mathrm{eV} \sum_{i,h,h_P} \left(\frac{n_\nu(\nu_{i,h})}{56\,\mathrm{cm}^{-3}}\right) A_{ii}\, \bigg[h_P \sin^2 2\theta_i \cos\Delta\Phi_i \, \left(2-\bar{\beta}_i^2\right)  \\
    &\qquad\qquad\qquad\qquad\qquad\qquad+ h\left(\frac{\cos^4\theta_i }{\beta_{i}^-}\left(3-(\beta_{i}^-)^2\right) +\frac{\sin^4\theta_i }{\beta_{i}^+}\left(3-(\beta_{i}^+)^2\right)\right) \bigg].
    \end{split}
\end{equation}
 {Notice how the last term in~\eqref{eq:stodSplitting} appears to diverge when the limit $\beta_{i}^\pm \to 0$ is taken.} As discussed in~\cite{Bauer:2022lri}, this is a result of the frame transformation, and so must cancel against some other term to leave the final result finite. The relevant term here is $n_{\nu}(\nu_{i,h}) h$, which due to the relative motion of the Earth vanishes when $\beta_i^\pm \leq \beta_\Earth$ to leave the energy splitting finite.

The dimensionless kinematic terms being summed over have a maximum of $\mathcal{O}(10^{3})$ when $\beta_{i} \sim \mathcal{O}(10^{-3})$, which is close to the expected speed of relic neutrinos at $T_{\nu} = T_{\nu,0}$. We therefore expect energy splittings $\Delta E \sim \mathcal{O}(10^{-36})\,\mathrm{eV}$ due to the Stodolsky effect, assuming a maximal asymmetry between left- and right-helicity neutrinos. In standard cosmology, however, we do not expect any asymmetry. {  A recent work~\cite{Arvanitaki:2022oby} has suggested that a sizeable local asymmetry of $\mathcal{O}(10^{-4})$ may develop at late times due to effects at the surface of the Earth. However, several works~\cite{Kalia:2024xeq,Gruzinov:2024ciz,Huang:2024tog} have since demonstrated that this effect is significantly smaller, of $\mathcal{O}(10^{-8})$,
when the spherical geometry of the Earth is taken into account.} On the other hand, constraints from Big Bang nucleosynthesis also allow for early universe asymmetries of $\mathcal{O}(10^{-2})$~\cite{Mangano:2011ip, Aver:2015iza}, which may persist today. We will use this as our benchmark value when plotting the sensitivity to the Stodolsky effect.

As before, we can now examine the two limits. In the seesaw limit, only the light neutrino terms survive
\begin{equation}
    \Delta E_\mathrm{SS} = \frac{2\sqrt{2}G_F}{3} \beta_\Earth \sum_{i,h} n_{\nu}(\nu_{i,h})\,h \frac{ A_{ii}}{\beta_{i}^-}\left(3-(\beta_{i}^-)^2\right),
\end{equation}
recovering the Majorana neutrino result from~\cite{Bauer:2022lri}. We can also recover the Dirac limit by taking the Majorana mass to zero, in which case $\theta_i \to \pi/4$, $\Delta\Phi_i \to 0$, and all kinematic terms for the $\pm$ states become equal. This leaves
\begin{equation}\label{eq:diracSplitting}
    \Delta E_\mathrm{D} = \frac{\sqrt{2}G_F}{3} \beta_\Earth \sum_{i,h} n_{\nu}(\nu_{i,h}) A_{ii} \bigg[2h_P \left(2-\bar{\beta}_i^2\right) + \frac{h}{\beta_{i}^-}\left(3-(\bar{\beta}_{i})^2\right)\bigg].
\end{equation}
Comparing to~\cite{Bauer:2022lri}, we see that the first term in~\eqref{eq:diracSplitting} is the one typically proportional to the difference between the number of neutrinos and antineutrinos. Here, it is instead proportional to the difference between early-universe left- and right-helicity neutrinos, justifying our comparison of the linear combination with $h_P = -1$ to Dirac neutrinos, and those with $h_P = +1$ to Dirac antineutrinos.

The energy splitting~\eqref{eq:stodSplitting} is far too small to be observed using spectroscopic methods, but may instead be inferred by measuring effects resulting from spin precessions induced by the non-commuting spin operators and the now spin-dependent electron Hamiltonian. There are two such effects, the first is a time-dependent magnetisation of a target, transverse to the incident neutrino wind, which may be measured using a SQUID magnetometer. The second is the result of tiny torques acting on a polarised target, which give rise to a net acceleration, measurable with an extremely sensitive torsion balance. Existing techniques are expected to be sensitive to energy splittings of $\Delta E \simeq 10^{-32}\,\mathrm{eV}$ and $\Delta E \simeq 5.2\cdot 10^{-28}\,\mathrm{eV}$, for the two effects, respectively~\cite{Rostagni:2023eic}. However, a future torsion balance experiment utilising a torsion balance suspended by superconducting magnets~\cite{Hagmann:1999kf} could be sensitive to energy splittings as small as $\Delta E \sim \mathcal{O}(10^{-36})\,\mathrm{eV}$, in range of detecting the C$\nu$B assuming a maximal asymmetry. In what follows, we will use the more realistic SQUID magnetometer result of $\Delta E \simeq 10^{-32}\,\mathrm{eV}$ as our reference sensitivity. 

\subsection{Coherent scattering}\label{sec:coherent}
We now focus on coherent scattering, where a neutrino scatters elastically\footnote{At the extremely low energies involved in relic neutrino coherent scattering, there is also a quasi-elastic charged current contribution to the coherent scattering rate on electrons~\cite{Bauer:2022lri}.} on a target $X$ in the process
\begin{equation}
        \nu_{i} + X \to \nu_i + X,
\end{equation}
at very low energies, such that its wavelength, or more specifically the inverse of the momentum transfer, is approximately the size of the target. When this condition is satisfied, the contributions to the neutrino scattering amplitude from each target within one neutrino wavelength add coherently, such that total scattering scales with the number of targets squared, as opposed to the usual linear scaling with the number of targets. This process has already been observed by the COHERENT collaboration for neutrinos with energies $E_{i} \sim \mathcal{O}(\mathrm{MeV})$~\cite{COHERENT:2017ipa,COHERENT:2020iec,COHERENT:2021xmm}, corresponding to neutrino wavelengths comparable to the size of atomic nuclei. As a result, the scattering cross sections observed by COHERENT are proportional to the square of the number of nucleons in the target. The expected enhancement for relic neutrinos is far more extreme, as their low momenta, $p_{i} \sim \mathcal{O}(\mathrm{meV})$, corresponds to wavelengths $\sim \mathcal{O}(\mathrm{mm})$, such that they are capable of scattering on macroscopic targets coherently. This, in turn, leads to scattering cross sections enhanced by a factor of order the Avogadro number, $N_A = 6.02 \cdot 10^{23}\,\mathrm{mol}^{-1}$, compared to the equivalent incoherent scattering rate, significantly 
boosting detection prospects. The scattering rates resulting from this enhancement for relic neutrinos are typically $\sim\mathcal{O}(\mathrm{kHz})$~\cite{Shergold:2021evs}, and can potentially be observed using an extremely sensitive torsion balance~\cite{Domcke:2017aqj,Shergold:2021evs,Bauer:2021uyj}, which experiences a small acceleration in the cosmic neutrino wind.

Following~\cite{Shergold:2021evs}, the coherent scattering of relic neutrinos on a target of mass $M$ induces a tiny acceleration of order
\begin{equation}\label{eq:cohAccelerationGeneralSingle}
    a_i = \frac{1}{M} \Gamma_i \Delta p_i,
\end{equation}
per degree of freedom, where $\Delta p_i$ is the net momentum transfer to the bulk target per scattering event, and the scattering rate is given by
\begin{equation}
    \Gamma_i = N_T n_\nu(\nu_{i,h})\beta_i \sigma_{\mathrm{C},i}, 
\end{equation}
with $\sigma_{C,i}$ the coherent scattering cross section, which will include the coherent enhancement factor, and $N_T$ is the number of targets. It therefore follows that in the MESM, the acceleration due to coherent scattering will instead take the form
\begin{equation}\label{eq:cohAccelerationGeneral}
    a = \frac{1}{M} \sum_{i,h,h_P} n_{\nu}(\nu_{i,h}) \Tr[\widetilde \sigma_\mathrm{C} \rho(h_P,z)], \qquad \widetilde \sigma_\mathrm{C} = \beta \sigma_\mathrm{C} \Delta p,
\end{equation}
The matrix associated with the generalised coherent scattering cross section, weighted by the net momentum transfer, is given by
\begin{equation}\label{eq:coherentMatrix}
    \begin{split}
        \widetilde \sigma_\mathrm{C} &= \begin{pmatrix}
            \widetilde\sigma_\mathrm{C}^{--} & \widetilde\sigma_\mathrm{C}^{-+} \\
            \widetilde\sigma_\mathrm{C}^{+-} & \widetilde\sigma_\mathrm{C}^{++}
        \end{pmatrix} \\
        &= 2\pi^2 G_F^2\frac{N_A}{A m_A}\rho\begin{pmatrix}
            \cos^2\theta_i \beta_i^- \Delta p_i^- \left(\Upsilon^-_1 - \Upsilon^-_2\right) &\frac{i}{2}\sin 2\theta_i \bar \beta_i \Delta \bar p_i\Upsilon_3 \\
            -\frac{i}{2}\sin 2\theta_i \bar \beta_i \Delta \bar p_i \Upsilon_3 &  \sin^2\theta_i \beta_i^+ \Delta p_i^+ \left(\Upsilon^+_1 + \Upsilon^+_2\right)
        \end{pmatrix},
    \end{split}
\end{equation}
where we have used the shorthand
\begin{align}
    \Upsilon_1^x &= N_\mathrm{C}^{xx} (E^x_i)^2 (Q_V^2 + 3Q_A^2) , \\
    \Upsilon_2^x &= m_i^x (Q_V^2 - 3Q_A^2)(N_\mathrm{C}^{x-}\cos^2\theta_i m^-_i - N_\mathrm{C}^{x+}\sin^2\theta_i m^+_i), \\
    \Upsilon_3 &= \bar{N}_\mathrm{C}h \bar{E}_i^2 \bar{\beta}_i (Q_V^2 + 3Q_A^2).
\end{align}
with $Q_V \simeq A - Z$ and $Q_A = A - 2Z$ the vector and axial couplings, respectively, given in terms of the mass number, $A$, and atomic number of the target, $Z$, and where
\begin{equation}\label{eq:coherenceFactor}
    N_\mathrm{C}^{xy} = \mathcal{H}\left(E_i^x - m_i^y - \frac{(m_i^y)^2 - (m_{i}^x)^2}{2M_N}\right) \left(\frac{1}{p_i^x}\right)^3,
\end{equation}
is the normalised\footnote{We extract the eigenstate-independent factor $(2\pi)^3 \frac{N_A}{A m_A}\rho$ that forms part of the prefactor in~\eqref{eq:coherentMatrix}.} coherent enhancement factor, which accounts for the increased cross section due to small momentum transfer in a typical relic neutrino scattering event. The step function here accounts for the fact that the $\nu_i^- \to \nu_i^+$ process can only take place if the $\nu_i^-$ has sufficient energy to scatter into the heavier $\nu_i^+$ state. Here we have also introduced the ``Avogadro mass'' $m_A \simeq 1\,\mathrm{g}\,\mathrm{mol}^{-1}$, and the mass density of the target, $\rho$. The step function here accounts for the fact that the $\nu_i^- \to \nu_i^+$ process can only take place if the $\nu_i^-$ has sufficient energy to scatter into the heavier $\nu_i^+$ state. Here we have also introduced the ``Avogadro mass'' $m_A \simeq 1\,\mathrm{g}\,\mathrm{mol}^{-1}$, and the mass density of the target, $\rho$.

{  More precisely, the denominator of~\eqref{eq:coherenceFactor} should contain the modulus of the 3-momentum transfer, $q$, in place of $p_i^x$. We will demonstrate, however, that setting $q \simeq p_i^x$ is always justified. Conserving energy and momentum in the most conservative scenario where the final state nucleus recoils antiparallel to the outgoing neutrino, we have
\begin{align}
    (p_i^x)^2 &\simeq (p_i^y)^2 + q^2 - 2 p_i^y q, \\
    E_i^x &\simeq E_i^y + \frac{q^2}{2M_N},
\end{align}
where $x$ and $y$ denote the initial and final mass eigenstates involved in the scattering process, respectively. To leading order in small parameters, $m_i^{x,y}/M$, $p_i^{x,y}/M$, and $q/M$, this has solution
\begin{equation}
    q^2 = (p_i^{x})^2 + \delta m_i^{yx} + \sqrt{(p_i^{x})^2 - \delta m_i^{yx}}\sqrt{(p_i^{x})^2 + 3\delta m_i^{yx}},
\end{equation}
with $\delta m_i^{yx} = (m_i^{y})^2 - (m_i^{x})^2$. Using this result, we immediately see that for mass-diagonal scattering, $q = p_i^x$ in both the pseudo-Dirac and seesaw limits. For non-diagonal scattering, we will focus on the two limits in turn. In the pseudo-Dirac limit, $\delta m_i^{yx} \ll (p_i^{x})^2$ and we can safely set $q \simeq p_{i}^{x}$. In the seesaw limit, $\delta m_i^{yx} \gg (p_i^{x})^2$. However, all terms contributing to off-diagonal scattering are proportional to $\sin\theta_i$, which vanishes in the seesaw limit. Thus we can safely set $q = p_i^{x}$ in all cases.
}

Taking the trace of the generalised cross section multiplied by the density matrix, we find
\begin{equation}
    \begin{split}
     \Tr[\sigma_{\rm C}\, \rho_i(h_P,z)] &= 2\pi^2 G_F^2 \frac{N_A}{A m_A}\rho\bigg[\cos^4\theta_i \beta_i^- \Delta p_i^- (\Upsilon_1^- - \Upsilon_2^-)\\ 
     &+ \sin^4\theta_i \beta_i^+ \Delta p_i^+ (\Upsilon_1^+ + \Upsilon_2^+) + \frac{h_P}{2} \sin^2 2\theta_i \bar\beta_i \Delta \bar p_i \cos\Delta\Phi_i \Upsilon_3\bigg].
    \end{split}
\end{equation}
Finally, inserting the trace into~\eqref{eq:cohAccelerationGeneral}, and simplifying the overall prefactor, we find the acceleration due to coherent scattering in the MESM
\begin{equation}\label{eq:coherentAcc} 
    \begin{split}
    a &= 2\pi^2 G_F^2\left(\frac{N_A}{A m_A}\right)^2 \rho \sum_{i,h,h_P}n_{\nu}(\nu_{i,h})  \bigg[\cos^4\theta_i \beta_i^-\Delta p_i^- (\Upsilon_1^- - \Upsilon_2^-) \\
    &\qquad\qquad\qquad\qquad\qquad + \sin^4\theta_i \beta_i^+\Delta p_i^+(\Upsilon_1^+ + \Upsilon_2^+) + \frac{h_P}{2} \sin^2 2\theta_i \bar\beta_i\Delta \bar p_i \cos\Delta\Phi_i \Upsilon_3\bigg] \\
    &= 6.16 \cdot 10^{-34}\,\mathrm{cm}\,\mathrm{s}^{-2} \sum_{i,h,h_P}\left(\frac{n_\nu(\nu_{i,h})}{56\,\mathrm{cm}^{-3}}\right) \bigg[\cos^4\theta_i \beta_i^- \Delta p_i^- (\Upsilon_1^- - \Upsilon_2^-)  \\
    &\qquad\qquad\qquad\qquad\qquad  + \sin^4\theta_i \beta_i^+ \Delta p_i^+ (\Upsilon_1^+ + \Upsilon_2^+) + \frac{h_P}{2} \sin^2 2\theta_i \bar\beta_i \Delta \bar p_i\cos\Delta\Phi_i \Upsilon_3\bigg].
    \end{split}
\end{equation}
where we have used the mass density $\rho = 7.87\,\mathrm{g}\,\mathrm{cm}^{-3}$ for ${^{56}\mathrm{Fe}}$, for which $Q_V \simeq 28$ and $Q_A = 4$, and the average bulk momentum transferred to the target by each scattering event is
\begin{equation}
    \Delta p_{i}^{x} \simeq \frac{\beta_\Earth}{3E_{i}^x}\left(4(E_{i}^x)^2 - (p_{i}^x)^2 \right),
\end{equation}
which takes into account the corrections to isotropic scattering due to the movement of the Earth through the C$\nu$B.

As before, we would now like to check the behaviour of the acceleration in the two limits, to ensure that it reproduces the results for Dirac and seesaw Majorana neutrinos as given in~\cite{Bauer:2022lri}. First, we check the seesaw limit, found by taking $\theta_i \to 0$. This leaves
\begin{equation}
    \begin{split}
    a_\mathrm{SS} &= 2\pi^2 G_F^2\left(\frac{N_A}{A m_A}\right)^2 \rho \sum_{i,h,h_P}n_{\nu}(\nu_{i,h})  \frac{\beta_i^-\Delta p_i^-}{(p_i^-)^3} \big[(E_i^-)^2(Q_V^2 + 3Q_A^2) \\
    &\qquad\qquad\qquad\qquad\qquad\qquad\qquad\qquad\qquad\qquad\qquad\qquad- (m_i^-)^2 (Q_V^2 - 3Q_A^2)\big] \\
    &= 2\pi^2 G_F^2\left(\frac{N_A}{A m_A}\right)^2 \rho \sum_{i,h,h_P}n_{\nu}(\nu_{i,h}) \frac{ E_i^- \Delta p_i^-}{(p_i^-)^2} \left((\beta_i^-)^2 Q_V^2 + 2(3-(\beta_i^-)^2)Q_A^2\right),
    \end{split}
\end{equation}
where in going from the first to second line we have expressed the mass of the final state neutrino in terms of the initial state neutrino energy. This exactly reproduces the seesaw limit given in~\cite{Bauer:2022lri}. In the Dirac limit, where $\theta_i \to \pi/4$ and $m_i^+ = m_i^- = \bar{m}_i$, we instead recover
\begin{equation}
    a_\mathrm{D} = \pi^2 G_F^2\left(\frac{N_A}{A m_A}\right)^2 (Q_V^2 + 3Q_A^2) \rho \sum_{i,h,h_P}n_{\nu}(\nu_{i,h}) \frac{ \bar E_i \Delta \bar p_i}{\bar p_i^2}(1+h h_P \bar\beta_i),
\end{equation}
which again recovers the results of~\cite{Bauer:2022lri} for Dirac neutrinos ($h_P = -1$) and antineutrinos ($h_P = +1$).

For our reference sensitivity, we use the existing torsion balances discussed in Section~\ref{sec:stodolsky}, which are capable of measuring accelerations as small as $a = 10^{-15}\,\mathrm{cm}\,\mathrm{s}^{-2}$, but note that a future torsion balance suspended by superconducting magnets may be capable of measuring accelerations as small as $a = 10^{-23}\,\mathrm{cm}\,\mathrm{s}^{-2}$~\cite{Hagmann:1999kf}.

\subsection{Accelerator}\label{sec:accelerator}
The last proposal that we will consider is an accelerator experiment~\cite{Bauer:2021uyj}, where a beam of highly charged ions is accelerated through the C$\nu$B such that relic neutrinos interact resonantly in one of the processes
\begin{align}
    &\nu_e + {^{A}_Z P} \to e^-\,(\mathrm{bound}) + {^{A}_{Z+1} D}, \\
    &\bar\nu_e + e^-\,(\mathrm{bound}) + {^{A}_Z P} \to {^{A}_{Z-1} D},
\end{align}
which we refer to as resonant bound beta decay (RB$\beta$) and resonant electron capture (REC), respectively. Here, $P$ and $D$ refer in turn to the parent and daughter ions. The observable in this proposal is the number of daughter ions on the beam after a given runtime. These can be counted by first fully ionising the beam, and then separating parent and daughter ions by their charge-mass ratio using a magnetic field. Directly on resonance, relic neutrinos have a large cross section to interact with the beam ions, with cross sections as large as $\sim \mathcal{O}(10^{-15}\,\mathrm{cm}^2)$, exceeding even those of electromagnetic processes. This is balanced by both the low number of beam ions that can be used at any one time without causing damage to the experimental apparatus, as compared to the fixed target experiments discussed so far, and the extreme beam energies required to hit such a resonance.

Written in terms of the normalised experimental runtime, $x = \frac{t}{\gamma_b \tau_D}$, where $t$ is the actual runtime, $\gamma_b$ is the Lorentz factor of the ion beam, and $\tau_D$ is the daughter state lifetime, the expected number of daughter states on the beam is given by~\cite{Bauer:2022lri}
\begin{equation}
    \begin{split}
    N_D(x) = N_P (1-e^{-x}) \sum_{i,h,h_P} n_{\nu}(\nu_{i,h})\Tr[R_\tau \rho(h_P,z)] + \mathcal{O}(R_\tau),
    \end{split}
\end{equation}
where $N_P$ is the number of parent states on the beam at $t = 0$, and the dimensionless quality factor $R_\tau$ is the ratio of the neutrino capture rate to the effective daughter lifetime, which we have normalised by the neutrino density. In the narrow resonance limit, the quality factor in the MESM reads
\begin{equation}
    \begin{split}
    R_\tau &= \begin{pmatrix}
    R_\tau^{--} & R_\tau^{-+} \\
    R_\tau^{+-} & R_\tau^{++}
    \end{pmatrix}\\
    &= |U_{ei}|^2\sqrt{\frac{\pi^3}{2}}\left(\frac{2J_D + 1}{2J_P+1}\right)\frac{\gamma_b}{Q^2}\mathcal{B}_{DP} \begin{pmatrix}
        \cos^2\theta_i f_i^- & \pm \frac{i}{2}\sin 2\theta_i \bar f_i\\
        \mp \frac{i}{2} \sin 2\theta_i \bar f_i & \sin^2\theta_i f^+_i
    \end{pmatrix},
    \end{split}
\end{equation}
where the upper signs should be taken for RB$\beta$ processes, since there is a neutrino in the initial state, and the lower signs should be taken for REC processes, where there is instead an antineutrino in the initial state. Here, $J_D$ and $J_P$ are the spins of the daughter and final state ions, respectively, $Q$ is the threshold for neutrino capture, and $\mathcal{B}_{DP}$ is the branching fraction for the daughter state to decay back to the parent state, which we assume to be approximately independent of the neutrino mass. The beam rest frame neutrino distributions take the form
\begin{equation}
    f_i^x \simeq \frac{1}{\mu_i^x \sqrt{(\delta_{i}^x)^2 + \delta_b^2}}\exp\left[-\frac{1}{2}\left(\frac{Q-\mu_{i}^x}{\mu_{i}^x\sqrt{(\delta_{i}^x)^2 + \delta_b^2}}\right)^2\right]
\end{equation}
where $\delta_b$ and $\delta_{i}^x$ are the normalised lab frame beam and neutrino momenta spreads, respectively, and $\mu_{i}^x$ is the mean beam frame energy of the relic neutrinos.
The maximum number of parent states that can put on the beam before damaging the experiment for an LHC-like circular accelerator has been estimated in~\cite{Bauer:2021uyj}, and is given by
\begin{equation}
    N_P = \frac{\mathcal{N}}{I^2\gamma_b^5}, \qquad \mathcal{N} \simeq 9.68\cdot 10^{40},
\end{equation}
where $I \simeq Z$ is the ionisation of the parent states on the beam, approximately equal to their atomic number. In the ideal setup, we would $\mu_{i}^x = Q$ for one of the neutrino states, putting it directly on resonance, whilst the others satisfy $\mu_{j}^y = E_{j}^y Q/E_{i}^x$. Denoting the energy of the neutrino that is directly on resonance as $E_{\nu,\mathrm{res}}$, we can therefore rewrite the Lorentz factor and beam rest frame distributions as
\begin{equation}
    \gamma_b = \frac{Q}{E_{\nu,\mathrm{res}}}, \qquad \mathcal{F}_i^x = \Delta m \gamma_b f_i^x = \frac{\Delta m}{E_i^x \sqrt{(\delta_{i}^x)^2 + \delta_b^2}} \exp\left[-\frac{1}{2}\left(\frac{E_{\nu,\mathrm{res}} - E_i^x}{E_{i}^x\sqrt{(\delta_{i}^x)^2 + \delta_b^2}}\right)^2\right],
\end{equation}
where $\Delta m = 50\,\mathrm{meV} \simeq \sqrt{|\Delta m_{31}^2|}$ is reference energy scale used for normalisation, which will be of a similar order to the heaviest, light neutrino mass. Collecting everything together, we therefore find for the total number of ions on the beam at normalised runtime $x$,
\begin{equation}\label{eq:nDaughter}
    \begin{split}
    N_D(x) &= \mathcal{N}\sqrt{\frac{\pi^3}{2}}\left(\frac{2J_D + 1}{2J_P+1}\right)\frac{E_{\nu,\mathrm{res}}^5}{I^2 Q^7 \Delta m}\mathcal{B}_{DP}(1-e^{-x})\sum_{i,h,h_P}|U_{ei}|^2 n_{\nu}(\nu_{i,h})\\
    &\qquad\qquad\qquad\qquad\qquad\times\left[\cos^4\theta_i \mathcal{F}_i^- + \sin^4\theta_i \mathcal{F}_i^+ \mp \frac{h_P}{2} \sin^2 2\theta_i \cos\Delta \Phi_i \bar{\mathcal{F}}_i\right].
    \end{split}
\end{equation}
Before considering a specific system, we will first examine the behaviour of the term in square brackets in the two limits. In the seesaw limit, we are left with
\begin{equation}
    \begin{split}
    N_{D,\mathrm{SS}}(x) = \mathcal{N} \sqrt{\frac{\pi^3}{2}}\left(\frac{2J_D + 1}{2J_P+1}\right)\frac{E_{\nu,\mathrm{res}}^5}{I^2 Q^7}\mathcal{B}_{DP}&(1-e^{-x})\sum_{i,h,h_P} \frac{|U_{ei}|^2 n_{\nu}(\nu_{i,h})}{E_{i}^-\sqrt{(\delta_{i}^-)^2 + \delta_b^2}}\\
    &\times\exp\left[-\frac{1}{2}\left(\frac{E_{\nu,\mathrm{res}}-E_{i}^-}{E_{i}^-\sqrt{(\delta_{i}^-)^2 + \delta_b^2}}\right)^2\right],
    \end{split}
\end{equation}
precisely the result of~\cite{Bauer:2022lri}. In this limit, both the left and right production helicity states contribute equally, giving a rate that is twice as large as the Dirac neutrino rate, where only one of the fluxes contributes. For Dirac neutrinos, this is the neutrino flux in the RB$\beta$ process, analogous to the $h_P = -1$ state, and the antineutrino flux in the REC process, analogous to $h_P = +1$. This is clear from the pseudo-Dirac limit, where $\mathcal{F}_i^- \simeq \mathcal{F}_i^+ \simeq \bar{\mathcal{F}}$, such that
\begin{equation}
    \begin{split}
    N_{D,\mathrm{PD}}(x) = \mathcal{N} \sqrt{\frac{\pi^3}{2}}&\left(\frac{2J_D + 1}{2J_P+1}\right)\frac{E_{\nu,\mathrm{res}}^5}{I^2 Q^7}\mathcal{B}_{DP}(1-e^{-x})\sum_{i,h,h_P} \frac{|U_{ei}|^2 n_{\nu}(\nu_{i,h})}{\bar E_{i}\sqrt{\bar\delta_{i}^2 + \delta_b^2}}\\
    &\times\exp\left[-\frac{1}{2}\left(\frac{E_{\nu,\mathrm{res}}-\bar E_{i}}{\bar E_{i}\sqrt{\bar\delta_{i}^2 + \delta_b^2}}\right)^2\right]\left[1-\sin^2 2\theta_i (1\pm h_P \cos \Delta\Phi_i)\right],
    \end{split}
\end{equation}
where we remind the reader that the upper sign corresponds to the RB$\beta$ process, and the lower sign to the REC process. The final term in square brackets is nothing more than the probability of oscillating to a detectable state given in~\eqref{eq:probs_ptol} for the RB$\beta$ process, and its analogue for `antineutrinos' for the REC process. This, in turn, selects just one of the production states in the Dirac limit, $\Delta \Phi_i \to 0$ and $\theta_i \to \pi/4$, as expected.

Given the scaling with $E_{\nu,\mathrm{res}}$, the efficacy of this proposal is clearly maximised when the beam energy is chosen such that heaviest active neutrino state is directly on resonance, that is, when $E_{\nu,\mathrm{res}} = E_{\mathrm{heavy}}$. However, the heaviest neutrino state (either $\nu_{3}^+$ in the normal mass hierarchy, or $\nu_{2}^+$ in the inverted hierarchy) approximately aligns with the corresponding sterile neutrino state in the seesaw limit, such that the active neutrinos will be far from resonance. We should therefore choose the beam energy such that the heaviest, light neutrino state ($\nu_{3}^-$, or $\nu_{2}^-$) is on resonance. Given that the light and heavy neutrino eigenstates have approximately the same mass in the pseudo-Dirac limit, this will do little to diminish the sensitivity there, but will ensure that the capture rate does not drop to zero in the seesaw limit. We therefore choose the beam energy such that the resonance is centred on neutrinos with energy
\begin{equation}
    E_{\nu,\mathrm{res}} = \sqrt{(m_\mathrm{heavy}^{-})^2 + \left\langle p_{i}\right\rangle^2},
\end{equation}
where $\left\langle p_{i}\right\rangle = 3.15\,T_{\nu,0}$ is the mean momentum of relic neutrinos today. {  This corresponds to a beam energy per nucleon requirement of
\begin{equation}
        \frac{E_\mathrm{beam}}{A} \simeq \frac{(m_\mathrm{ion}/A)}{E_{\nu,\mathrm{res}}} Q = 10\,\mathrm{TeV}\,\left(\frac{(m_\mathrm{ion}/A)}{1\,\mathrm{GeV}}\right)\left(\frac{0.1\,\mathrm{eV}}{E_{\nu,\mathrm{res}}}\right)\left(\frac{Q}{1\,\mathrm{keV}}\right),
\end{equation}
where $m_\mathrm{ion}$ and $A$ denote the mass and mass number of the beam ion, respectively.}

Finally, we focus on the ${^{157}\mathrm{Gd}} \to {^{157}\mathrm{Tb}}$ system given in~\cite{Bauer:2021uyj}, for which $Q = 10.95\,\mathrm{keV}$, and find the expected number of daughter ions on the beam after normalised runtime $x$
\begin{equation}\label{eq:nDaughterNumeric}
    \begin{split}
    N_D(x) &= 1.38\cdot 10^{-8}\left(\frac{E_\mathrm{\nu,res}}{\Delta m}\right)^5(1-e^{-x})\sum_{i,h,h_P}|U_{ei}|^2 n_{\nu}(\nu_{i,h})\\
    &\qquad\qquad\qquad\qquad\qquad\times\left[\cos^4\theta_i \mathcal{F}_i^- + \sin^4\theta_i \mathcal{F}_i^+ + \frac{h_P}{2} \sin^2 2\theta_i \cos\Delta \Phi_i \bar{\mathcal{F}}_i\right].
    \end{split}
\end{equation}
where the normalised runtime scales as
\begin{equation}
    x = 4.46 \cdot 10^{-8} \left(\frac{E_{\nu,\mathrm{res}}}{\Delta m}\right)\left(\frac{t}{1\,\mathrm{y}}\right).
\end{equation}
In order to detect the C$\nu$B, we would require $N_P \sim \mathcal{O}(1)$ after a runtime $t\sim \mathcal{O}(\mathrm{y})$, which seems impossible with this setup. { Additionally, such an experiment would require a beam energy per nucleon of $\mathcal{O}(100\,\mathrm{TeV})$, assuming that the heaviest active neutrino mass scale is $\mathcal{O}(0.1\,\mathrm{eV})$.} We note, however, the huge $Q^{-7}$ scaling appearing in~\eqref{eq:nDaughter}, which would drastically improve detection prospects if a low enough $Q$ target was discovered. For example, a ${^{157}\mathrm{Gd}} \to {^{157}\mathrm{Tb}}$-like system with a threshold $Q \sim \mathcal{O}(0.1\, \mathrm{keV})$ could have sensitivity to the C$\nu$B after a few years of runtime. { This analogous system with a smaller threshold has the additional benefit of reducing the beam energy requirements to $\mathcal{O}(1\,\mathrm{TeV})$, significantly boosting the feasibility of this approach}. As discussed in~\cite{Bauer:2021uyj}, it is possible to artificially reduce the threshold by using excited ions on the beam; such systems can have a threshold $Q \sim \mathcal{O}(\mathrm{keV})$, however, these systems also come with additional experimental considerations and difficulties. For this reason, we will use the reference values in~\eqref{eq:nDaughterNumeric} in what follows.

\section{Results}\label{sec:results}

We now show the sensitivity of each of the C$\nu$B detection proposals discussed in Section~\ref{sec:detectors} to relic neutrinos in the MESM. In all cases, we plot the sensitivity, which we define as the momentum-averaged ratio
\begin{equation}
    \mathcal{S} = \int_0^\infty dp_{i}\, p_{i}^2 \bar{f}(p_{i})  \left(\frac{\mathcal{O}}{\mathcal{O}_0}\right),
\end{equation}
where $\mathcal{O}$ is the observable given in the corresponding subsection for each proposal, $\mathcal{O}_0$ is the reference sensitivity. We compute all sensitivities and oscillation probabilities on a $256\times 256$ log-uniform grid of $m_D \in [10^{-10}, 10]\,\mathrm{eV}$, and $m_R \in [10^{-35}, 10^{10}]\,\mathrm{eV}$, which is sufficient to show the transition from the seesaw to the pseudo-Dirac limit, the exclusion bounds from KATRIN and oscillation experiments, as well as the interesting behaviour when the oscillation baseline approaches the size of the observable universe. At each mass point, we average the sensitivity over $257$ uniformly spaced momentum points $p_{i} \in [0,12]\,\mathrm{T_{\nu,0}}$. This momentum range includes $\gtrsim 99.9\%$ of all neutrinos described by a Fermi-Dirac distribution at a temperature $T_{\nu,0}$.
\begin{table}[]
    \renewcommand{\arraystretch}{1.5}
    \centering
    \begin{tabular}{c|c|c}
        Proposal & Observable & $\mathcal{O}_0$ \\
        \hline\hline
        PTOLEMY & \eqref{eq:ptolemyRate}, $t = 5\,\mathrm{y}$ & $9\,\mathrm{events}$ \\
        \hline
        Stodolsky & \eqref{eq:stodSplitting} & $10^{-32}\,\mathrm{eV}$ \\
        \hline
        Coherent & \eqref{eq:coherentAcc} & $10^{-15}\,\mathrm{cm}\,\mathrm{s}^{-2}$ \\
        \hline
        Accelerator & \eqref{eq:nDaughterNumeric}, $t = 5\,\mathrm{y}$ & $9\,\mathrm{events}$
    \end{tabular}
    \caption{Observables and reference sensitivities, $\mathcal{O}_0$, used in Figure~\ref{fig:sensitivity_contours}. For the PTOLEMY and accelerator proposals we assume a runtime of $5\,\mathrm{y}$, and a minimum of $9$ events required for a $3\sigma$ discovery.}
    \label{tab:sensitivites}
\end{table}

\begin{figure}[]
    \centering
\begin{subfigure}{0.495\textwidth}
    \includegraphics[width=\textwidth]{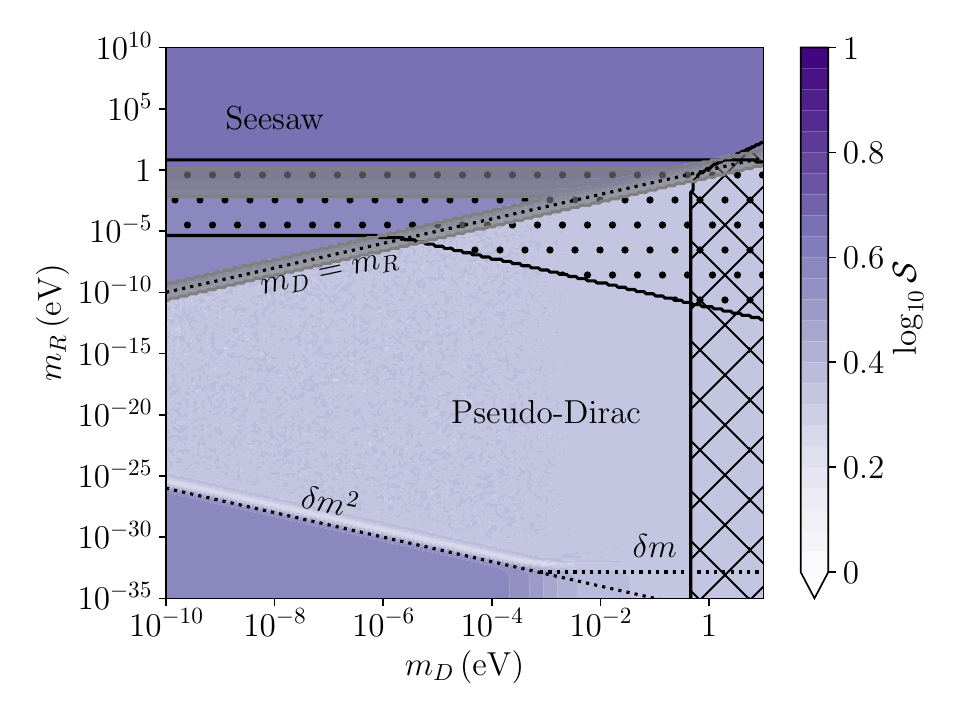}
    \caption{PTOLEMY}
    \label{fig:sensitivity_contours_ptolemy}
\end{subfigure}~
\begin{subfigure}{0.495\textwidth}
    \includegraphics[width=\textwidth]{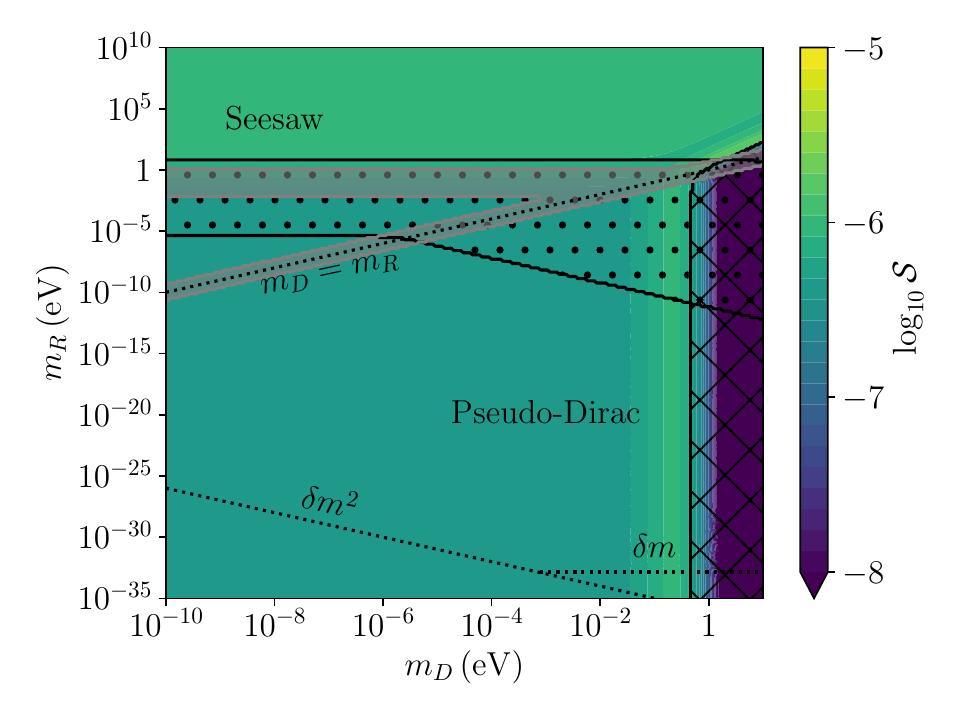}
    \caption{Stodolsky}
    \label{fig:sensitivity_contours_stodolsky}
\end{subfigure}
\begin{subfigure}{0.495\textwidth}
    \includegraphics[width=\textwidth]{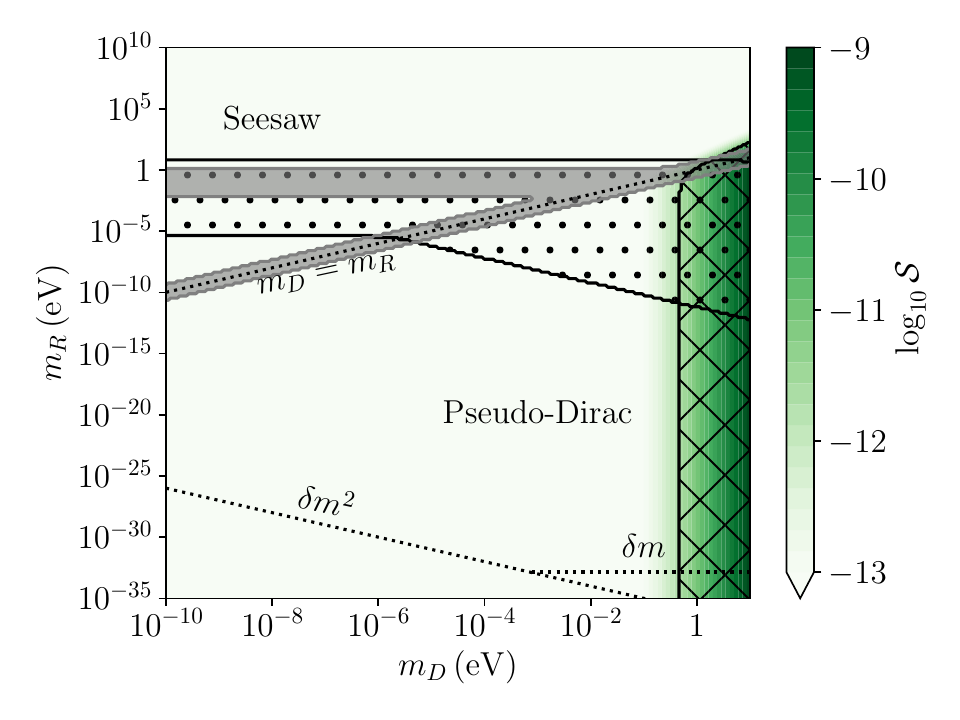}
    \caption{Coherent}
    \label{fig:sensitivity_contours_coherent}
\end{subfigure}~
\begin{subfigure}{0.495\textwidth}
    \includegraphics[width=\textwidth]{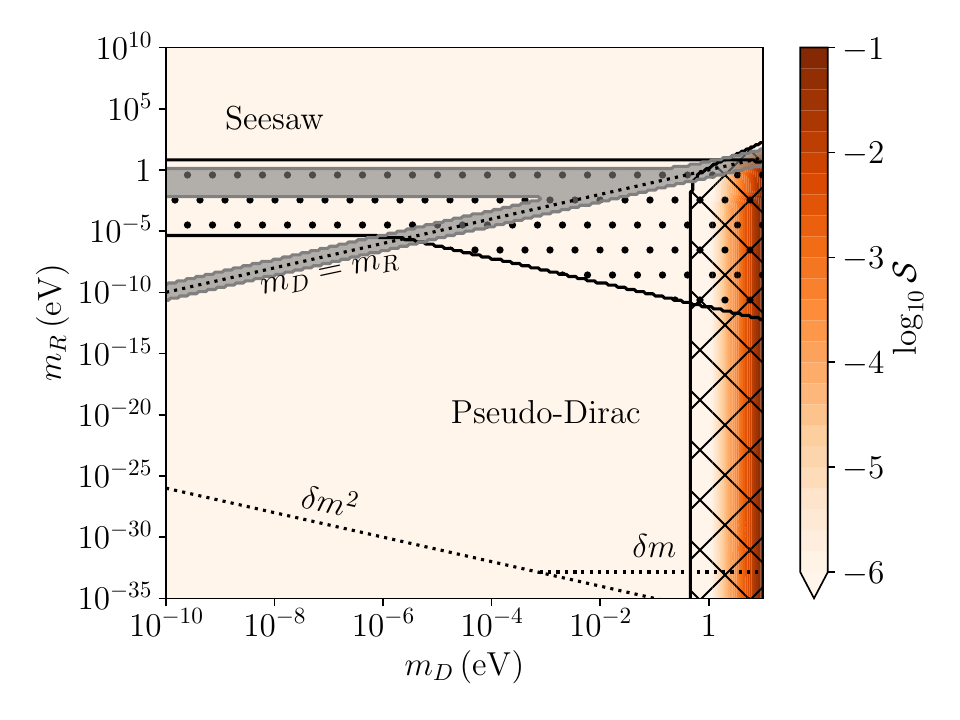}
    \caption{Accelerator}
    \label{fig:sensitivity_contours_accelerator}
\end{subfigure}    
    \caption{Sensitivity of a) the PTOLEMY proposal, b) the Stodolsky effect, c) coherent scattering, and d) an accelerator experiment to relic neutrinos in the MESM. The hatched regions are those excluded by the KATRIN experiment, whilst the dotted regions are those excluded by solar and atmospheric neutrino data. The grey regions highlight where the approximation in the computation of off-diagonal elements of the kinematic observable $K$ becomes inaccurate, explicitly where~\eqref{eq:uncertaintyApprox} is satisfied.}
    \label{fig:sensitivity_contours}
\end{figure}

We present the sensitivities for each of the four proposals in Figure~\ref{fig:sensitivity_contours}, which are computed according to the observables and reference sensitivities given in Table~\ref{tab:sensitivites}. 
As expected, the sensitivity of the PTOLEMY proposal, shown in Figure~\ref{fig:sensitivity_contours_ptolemy}, is almost exclusively dependent on how the mixing angles $\theta_i$ and the averaged cosine of the phase difference $\langle \cos\Delta\Phi_i\rangle$ vary depending on the parameters $m_R$ and $m_D$. 
The sensitivity is maximised in the seesaw limit, where the mixing angle tends to zero. 
In this limit, all six active neutrino states contribute with weights determined by the combination of the squared PMNS matrix element, $|U_{ei}|^2$, and the velocity factor, $1+\beta_{\nu_i}$ for left-helicity neutrinos, and $1-\beta_{\nu_i}$ for right-helicity neutrinos. 
In the normal mass hierarchy, the rate is therefore dominated by the lightest, left-helicity neutrino state, $\nu_{1,l}$, which has the largest PMNS matrix element, $|U_{e1}|^2 \simeq 0.68$, and picks up the enhancement factor $1+\beta_{\nu_1} \simeq 2$. 
 {For the design sensitivity of PTOLEMY, $\Delta \simeq 0.05\,\mathrm{eV}$~\cite{Apponi:2021hdu, Betti:2018bjv}, the contributions from each mass eigenstate will not be individually resolvable, instead appearing as a single peak in the electron energy spectrum.}
This behaviour is mirrored in the intermediate pseudo-Dirac regime, where the $\langle\cos\Delta\Phi_i\rangle$ tends instead towards zero. 
This, in turn, exactly halves the capture rate compared to the seesaw regime. 

The most interesting behaviour can be seen in the pseudo-Dirac regime when the inverse mass splittings approach the size of the observable universe, where the averaged cosine tends to one and the neutrinos do not evolve significantly over the age of the universe.
This implies that the right-helicity production states remain unobservable, such that only three left-helicity neutrino states contribute: two that are non-relativistic, weighted according to their small PMNS matrix elements, and the lightest state, dominating the rate due to its large PMNS matrix element, and whose contribution increases as it becomes more relativistic at small $m_D$. 
As a result, the rate in the small mass splitting, pseudo-Dirac, regime approaches the rate in the seesaw regime at small $m_D$, reaching a value of $\sim 82\%$ that of the seesaw value, confirming the results given in~\cite{Perez-Gonzalez:2023llw}.
Overall, the PTOLEMY proposal remains sensitive to the C$\nu$B, assuming the required energy resolution can be achieved, over the entire parameter space, with the event rate differing by at most a factor of two between the pseudo-Dirac and seesaw regions.

The sensitivity of a SQUID magnetometer to the Stodolsky effect is shown in Figure~\ref{fig:sensitivity_contours_stodolsky}. 
Unlike the PTOLEMY proposal, we find that the phase difference between the $\pm$ pair does not significantly alter the expected detection prospects for an experiment searching for the Stodolsky effect.
In fact, we find that the expected sensitivity is mostly affected by the neutrino mass spectrum.
When the neutrino masses are sufficiently large, the helicity asymmetry vanishes as $\beta_\Earth \geq \beta_i^\pm$, as discussed at length in~\cite{Bauer:2022lri}.
This occurs when both $m_D \gtrsim 0.5~{\rm eV}$ and $m_R \lesssim 10^{-1}~{\rm eV}$, and leads to a vanishing sensitivity to the Stodolsky effect, as seen in Figure~\ref{fig:sensitivity_contours_stodolsky}.
For values $m_D \sim 0.1~{\rm eV}$, the sensitivity is instead maximised due to the $1/\beta_{i}^\pm$ term in the energy splitting before it becomes significantly suppressed by the reduction of the helicity asymmetry. 
In the seesaw limit, we find that for values of $0.1~{\rm eV}\lesssim m_R \lesssim 10^5~{\rm eV}$, the aforementioned maximum of the sensitivity occurs for larger values of $m_D$, whilst the cancellation of the helicity asymmetry only occurs values of $m_D$ outside the range presented in this work.
This is expected in the seesaw limit, where the mass of the lightest neutrino is suppressed via the seesaw relation, $m_1^- \approx m_D^2/m_R$, whilst the other active neutrino masses tend to $m_2^- \to \sqrt{\Delta m_{21}^2}$ and $m_3^- \to \sqrt{\Delta m_{31}^2}$. As a result, neutrinos in this regime are either relativistic, or sufficiently warm that the relative motion of the Earth with respect to the C$\nu$B frame is insufficient to wash out  any helicity asymmetires.
 {Nevertheless, a successful detection of the C$\nu$B via the Stodolsky effect remains just beyond the reach of current technology, requiring a SQUID magnetometer with a sensitivity approximately six orders of magnitude higher than that of existing devices.}
Alternatively, the required sensitivity could be achieved if the torsion balance discussed in~\cite{Hagmann:1999kf} were realised.

In contrast to the PTOLEMY proposal, but similar to the Stodolsky effect, the coherent scattering and accelerator proposals depend more strongly on the neutrino kinematics than the phase difference and mixing. 
 {Both proposals strongly favour more massive neutrinos: in the case of coherent scattering, this preference arises from the larger momentum transfer per scattering event, while for the accelerator experiment, it stems from the lower beam energy requirement, which, in turn, allows for a greater number of parent ions in the beam without compromising the experimental setup.}
Unfortunately, the parameter region where these experiments are most sensitive is largely excluded by KATRIN.
As discussed in Section~\ref{sec:accelerator}, however, the accelerator proposal could become competitive if a target with a smaller threshold, $Q$, could be found. For example, a target with the same experimental parameters as the ${^{157}\mathrm{Gd}} \to {^{157}\mathrm{Tb}}$ system used in Figure~\ref{fig:sensitivity_contours_accelerator}, but with a threshold one order of magnitude smaller would be at least seven orders of magnitude more sensitive, or at least fourteen if two orders of magnitude can be found. This would make the accelerator proposal at least as sensitive as, if not more sensitive than, the PTOLEMY proposal to the C$\nu$B. Similarly, the torsion balance suspended by superconducting magnets proposed  {in}~\cite{Hagmann:1999kf} would increase the sensitivity of a coherent scattering experiment by approximately eight orders of magnitude, making it comparable to the Stodolsky effect without the requirement of a background asymmetry.

\section{Conclusions}\label{sec:conclusions}

The origin of neutrino masses remains an open question in high-energy physics. Given that the neutrino mass scale is significantly smaller than that of charged leptons, the most widely accepted explanation is the presence of very heavy right-chiral singlets, which suppress the active neutrino masses.
This suppression is intrinsically linked to the scale of lepton number violation through the Majorana masses carried by these right-chiral states. However, since the Majorana mass scale is not protected by Standard Model symmetries, it could be well below the electroweak scale.
Experimentally testing this scenario presents challenges. A small degree of lepton number violation induces a mass splitting between the two Majorana states that form a Dirac fermion. This mass difference could lead to long-distance oscillations, potentially affecting certain neutrino sources in nature.

In this work, we have explored how the introduction of a lepton number violating term in the Standard Model Lagrangian, via Majorana mass terms for additional right-chiral states, would influence various proposals for detecting the cosmic neutrino background.
Specifically, we have examined the impact of lepton number violation on neutrino capture by unstable nuclei, the Stodolsky effect, coherent scattering, and an accelerator experiment. In general, the sensitivities are found to smoothly transition between the previously established Dirac and Majorana limits reported in the literature.

Furthermore, the introduction of a Majorana mass term complicates the calculation of detection sensitivities. Since the initial C$\nu$B flux is generated in specific linear combinations, and both neutral and charged current interactions are non-diagonal in the extended mass basis, a density matrix formalism is necessary to accurately account for the potential interplay between propagation and scattering in a non-diagonal basis. Using the more conventional formalism based on oscillation probabilities would not reproduce the results found in the literature.

We have demonstrated that sensitivity of the detection proposals can be broadly divided into two main categories. The first are those whose sensitivity is predominantly influenced by the phase difference between the two mass eigenstates and their mixing, as is the case for the PTOLEMY proposal. The second category are those whose sensitivity is driven by the neutrino mass spectrum through kinematic effects. This applies to the Stodolsky effect, coherent scattering, and accelerator-based proposals.
For the first category, we observe a clear distinction between the pseudo-Dirac and seesaw limits. 
 {Within the pseudo-Dirac regime, there are two further sub-regions. When the mass splittings far exceed the inverse of the distance travelled by the C$\nu$B, the states produced in the early universe oscillate to different linear combinations in the present day. When the inverse is true, the present day C$\nu$B is populated by the same linear combinations as in the early universe. Depending on the experiment, this subtle distinction allows one to distinguish the two regimes, \textit{e.g.} at PTOLEMY, where the two regions have different sensitivity if neutrinos are relativistic.}
Additionally, we confirm that in the pseudo-Dirac scenario, the event rate is reduced to half of the expected rate in the seesaw scenario for a fully non-relativistic C$\nu$B, consistent with previous findings in the literature. For the second category, we find that the detection sensitivities are primarily affected by the neutrino mass spectrum. Specifically, in the case of the Stodolsky effect, the sensitivity tends to approach zero when the neutrino mass is large enough that the relative velocity of the Earth with respect to the C$\nu$B exceeds the individual neutrino velocities, resulting in the vanishing of the required helicity asymmetry.
The situation is reversed for coherent scattering and accelerator-based proposals. Since their sensitivities improve for more massive neutrinos, the effects of the lepton number violating terms are most significant when the Dirac masses are relatively large, $m_D \gtrsim 0.5~{\rm eV}$, and the Majorana masses are smaller, $m_R \lesssim 1~{\rm eV}$. This corresponds to the pseudo-Dirac region.
Unfortunately, however, the regions with the best sensitivities lie within the regions excluded by the KATRIN measurement.
 {When the Majorana mass is large, $m_R \gtrsim 1~{\rm eV}$, and the Dirac masses satisfy $m_D \lesssim m_R$, the seesaw mechanism suppresses the lightest neutrino mass, ensuring it remains relativistic even today.}
This, in turn, reduces the sensitivities for both the coherent scattering and accelerator-based detection proposals.

Given their differing sensitivities to a non-zero Majorana mass, we find that a positive signal at more than one of these C$\nu$B detection proposals would help to determine, or exclude, the scale of lepton number violation. However, achieving the necessary precision to observe these effects remains a significant technological challenge in all cases.
Nevertheless, we remain optimistic that in the not-too-distant future, the detection of the C$\nu$B will illuminate the origin of neutrino masses and reveal the scale of lepton number violation, should it exist, opening new doors in our understanding of fundamental physics.

%
%
\acknowledgments

Jack D. Shergold would like to thank the IPPP for their warm hospitality during the completion of the manuscript. YFPG was supported by the STFC under Grant No.~ST/T001011/1. YFPG also acknowledges financial support by the Consolidaci\'on Investigadora grant CNS2023-144536 from the Spanish Ministerio de Ciencia e Innovaci\'on (MCIN) and by the Spanish Research Agency (Agencia Estatal de Investigaci\'on) through the grant IFT Centro de Excelencia Severo Ochoa No CEX2020-001007-S. 
Jack D. Shergold is supported by the Spanish grants PID2023-147306NB-I00 and CEX2023-001292-S (MCIU/AEI/10.13039/501100011033), as well as CIPROM/2021/054 (Generalitat Valenciana).

\appendix
\section{Mixing factors}\label{app:mixingFactors}
When constructing amplitudes for the generalised cross sections, we need to ensure that mixing factors associated to each external particle are correctly included. To do so, we must carefully examine the terms in the Lagrangian that give rise to these mixing factors, and their interactions with the external states containing neutrinos. 

To begin, the amplitude should be constructed using the rules set out in~\cite{Gluza:1991wj}. There are then four kinds of terms that can appear in the generalised cross sections, arising from the combinations
\begin{equation}\label{eq:mixingTerms}
    (\Theta^x_i)^* \,\bar{\nu}_i^x \dots \psi \ket{\nu_i^x}, \qquad \Theta^x_i\, \bar\psi \dots \nu_i^x \ket{\nu_i^x}, \qquad (\Theta^x_i)^*\bra{\nu_i^x} \bar{\nu}_i^x \dots \psi, \qquad \Theta^x_i \bra{\nu_i^x} \bar\psi \dots \nu_i^x,
\end{equation}
where $\psi$ is some fermion field operator, and $\Theta^x_i$ is the mixing factor that comes from expanding out the $\nu_i$ field operators in terms of the $\pm$ states, with values
\begin{equation}
    \Theta_i^- = i \cos\theta_i,\qquad \Theta^+_i = \sin\theta_i.
\end{equation}
The four terms in~\eqref{eq:mixingTerms} are responsible for the spinors $\bar{v}_i^x$, $u_i^x$, $\bar{u}_i^x$, and $v_i^x$, respectively, such that in our amplitude we should assign a mixing factor to each spinor according to
\begin{equation}
    (\Theta_i^x)^* : \bar{v}_i^x, \bar{u}_i^x, \qquad \Theta_i^x : v_i^x, u_i^x.
\end{equation}
Any signs due to anticommutation relations appear as relative signs between diagrams, which should be taken care of in the usual manner.

%
%
\bibliographystyle{JHEP}
\bibliography{bibliography}

\end{document}